\documentclass[twocolumn]{aastex63}
\usepackage{psfig,enumerate}
\usepackage{color}
\usepackage{atbegshi}
\usepackage{url}

\usepackage[caption=false]{subfig}

\newcommand{\kms}{\mbox{km s$^{-1}~$}} 
\newcommand{\kmse}{\mbox{km s$^{-1}$}}

\newcommand{\vhelio}{$V_{\rm HELIO}~$}

\newcommand{\dgr}{$^{\circ}~$}

\newcommand{\h}{$^{\rm h}$}
\newcommand{\m}{$^{\rm m}$}
\newcommand{\s}{$^{\rm s}~$}

\newcommand{\gaia}{\textit{Gaia}}

\begin{document}

\title{Overview of The SDSS-V Magellanic Genesis Survey}
\correspondingauthor{David L. Nidever}
\email{dnidever@montana.edu}

\author[0000-0002-1793-3689]{David L. Nidever}
\affiliation{Department of Physics, Montana State University, P.O. Box 173840, Bozeman, MT 59717-3840}

\author[0000-0003-1856-2151]{Danny Horta}
\affiliation{Institute for Astronomy, University of Edinburgh, Royal Observatory, Edinburgh EH9 3HJ, UK}
\affiliation{Center for Computational Astrophysics, Flatiron Institute, 162 Fifth Avenue, New York, NY 10010, USA}

\author[0000-0003-2025-3147]{Steven R. Majewski}
\affiliation{Department of Astronomy, University of Virginia, Charlottesville, VA, 22904, USA}

\author[0009-0000-0733-2479]{Andres Almeida}
\affiliation{Department of Astronomy, University of Virginia, Charlottesville, VA, 22904, USA}

\author[0000-0002-6553-7082]{Joshua T. Povick}
\affiliation{Universidad Diego Portales, Santiago, Chile}

\author[0009-0000-9037-1697]{Slater J. Oden}
\affiliation{Department of Physics, Montana State University, P.O. Box 173840, Bozeman, MT 59717-3840}

\author[0000-0001-7434-5165]{\'Oscar Jim\'enez-Arranz}
\affiliation{Lund Observatory, Division of Astrophysics, Department of Physics, Lund University, Box 43, SE-22100, Lund, Sweden}

\author[0000-0003-1479-3059]{Guy Stringfellow}
\affiliation{Center for Astrophysics and Space Astronomy, University of Colorado, 389 UCB, Boulder, CO, 80309-0389, USA}

\author[0000-0001-9984-0891]{S. Drew Chojnowski}
\affiliation{NASA Ames Research Center, Moffett Field, CA 94035, USA}
\affiliation{Department of Physics, Montana State University, P.O. Box 173840, Bozeman, MT 59717-3840}

\author[0000-0001-7827-7825]{Roeland van der Marel}
\affiliation{Space Telescope Science Institute, 3700 San Martin Drive, Baltimore, MD 21218, USA}

\author[0000-0001-8536-0547]{Lara Cullinane}
\affiliation{Leibniz-Institut f\"ur Astrophysik (AIP), An der Sternwarte 16, D-14482 Potsdam, Germany}
\affiliation{The William H. Miller III Department of Physics \& Astronomy, Bloomberg Center for Physics and Astronomy, Johns Hopkins University, 3400 N. Charles Street, Baltimore, MD 21218, USA}

\author[0000-0003-4254-7111]{Bruno Dias}
\affiliation{Departamento de Fisica y Astronomia, Facultad de Ciencias Exactas, Universidad Andres Bello, Fernandez Concha
700, Las Condes, Santiago, Chile}

\author[0000-0001-7258-1834]{Jennifer Johnson}
\affiliation{Department of Astronomy and Center for Cosmology and AstroParticle Physics, The Ohio State University, Columbus, OH 43210,
USA}

\author[0009-0000-4049-5851]{John Donor}
\affiliation{Department of Physics and Astronomy, Texas Christian University, TCU Box 298840 Fort Worth, TX 76129, USA}

\author[0000-0002-6797-696X]{Maria-Rosa Cioni}
\affiliation{Leibniz-Institut f\"ur Astrophysik Potsdam, An der Sternwarte 16, D-14482 Potsdam, Germany}

\author[0000-0001-9852-1610]{Juna Kollmeier}
\affiliation{Observatories of the Carnegie Institution for Science, 813 Santa Barbara Street, Pasadena, CA, 91101, USA}

\author[0000-0003-0842-2374]{Andrew Tkachenko}
\affiliation{Institute of Astronomy, KU Leuven, Celestijnenlaan 200D, 3001 Leuven, Belgium}

\shorttitle{Magellanic Genesis Overview}
\shortauthors{Nidever et al.}

\begin{abstract}
The Sloan Digital Sky Survey-V (SDSS-V) Magellanic Genesis survey is a spectroscopic program designed to map the kinematic and chemical structure of the Magellanic Clouds using \textsl{APOGEE} and \textsl{BOSS} spectroscopy. This overview describes the survey's design, target selection, and science goals, and highlights some first results using these data. In the inner regions of the Large and Small Magellanic Clouds (LMC and SMC), the survey obtained high-resolution near-infrared \textsl{APOGEE} spectra (S/N$\approx$45) of $\sim$14{,}000 bright, oxygen-rich asymptotic giant branch (AGB-O) stars. These data provide contiguous spatial coverage of the Clouds' main bodies, enabling detailed chemo-dynamical studies.
To explore extended structures, the survey includes \textsl{BOSS} optical spectroscopy of fainter red giant (RG) stars selected with \gaia~DR3 data, reaching $G \approx 17.5$. Many of these targets extend to the outer regions of the Clouds, which are known to span $\sim$20$^\circ$ (LMC) and $\sim$12$^\circ$ (SMC) and contain diffuse substructures of unclear origin. \textsl{BOSS} data in the inner regions also complement \textsl{APOGEE} by providing elements inaccessible in the near-infrared and enabling cross-calibration between instruments.
The survey further includes \textsl{APOGEE} and \textsl{BOSS} observations of $\sim$300 evolved massive stars and a small sample of symbiotic binaries previously observed by \textsl{APOGEE}-1 and -2, enhancing our understanding of massive stellar evolution and complementing the SDSS-V main-sequence massive star program.
\end{abstract}


\section{Introduction}
\label{sec:intro}

The Large and Small Magellanic Clouds (LMC and SMC) are the most massive and prominent satellite galaxies of the Milky Way (MW). At distances of approximately 50 and 60 kpc \citep[][]{Pietrzynski2019,Cioni2000}, respectively, the Magellanic Clouds (MCs) are the MW's closest neighbors, offering a unique opportunity to study the structure, chemistry, kinematics, and evolution of dwarf galaxies in unprecedented detail. Their proximity enables high signal-to-noise ratio (S/N) observations of individual stars, and their status as interacting companions to each other and to the MW provides an astrophysical laboratory for understanding the dynamics of hierarchical galaxy formation, satellite interactions, and the assembly history of the MW halo.
\citep[e.g.,][]{vanDerMarel2004,Bekki2005,Besla2007,Besla2012,Jethwa2016,Nadler2020,Fox2016}.

Despite their significance, the MCs present observational challenges. Their large angular sizes on the sky --- spanning over 20$^\circ$ for the LMC and 12$^\circ$ for the SMC --- make it challenging to conduct homogeneous spectroscopic surveys. While deep and contiguous photometric surveys have made great strides in mapping the stellar populations of the MCs (e.g., SMASH; \citealt{Nidever2017,Nidever2020}, DELVE; \citealt{Drlica-Wagner2021}, MCPS; \citealt{Zaritsky2004}, OGLE; \citealt{Wyrzykowski2011}, STEP; \citealt{Ripepi_2014}, VISCACHA; \citealt{Maia2019}), and in deriving spatially resolved star formation histories (e.g., SMASH; \citealt{RuizLara2020,Massana2022}, and VMC; \citealt{Cioni2011, Rubele2018, Mazzi2021}, MagES; \citealt{Cullinane_2020}), spectroscopic coverage—particularly at high resolution and over wide areas is the natural next step.

\begin{figure}[t]
\begin{center}
\includegraphics[width=1.0\hsize,angle=0]{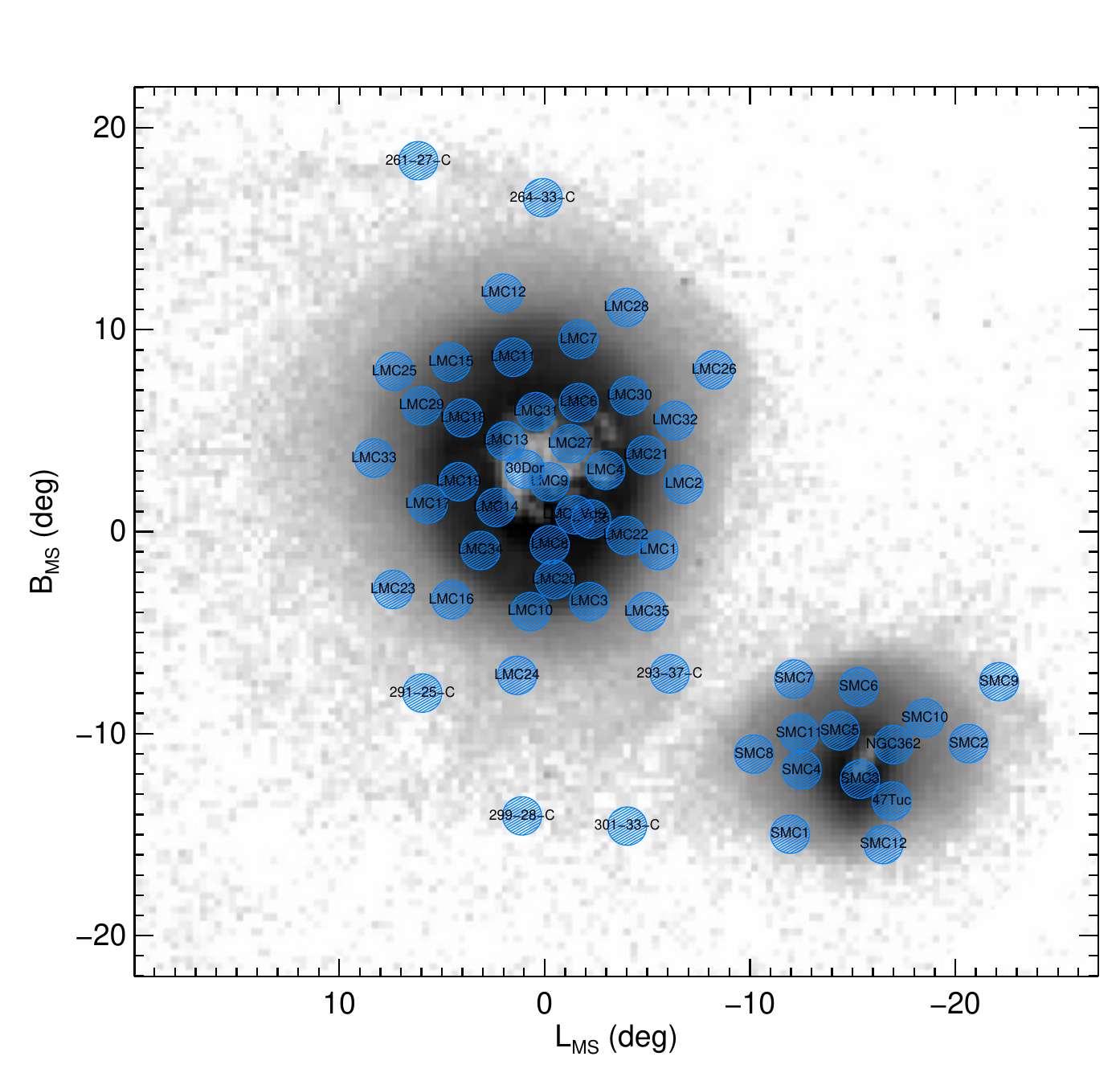}
\end{center}
\caption{Map of the SDSS-IV / \textsl{APOGEE}-2S MCs survey fields (blue) shown on top of the \citet{Belokurov2019} RGB star density map (grayscale), projected in the Magellanic Stream coordinate system \citep{Nidever2008}. While the SDSS-IV data sample a large radial and azimuthal range in the MCs, they only fill 33\% of the area in the LMC and SMC main bodies. Conversely, the Magellanic Genesis Survey covers the entire Magellanic system.}
\label{fig:apogee2map}
\end{figure}


\begin{figure*}[t]
\begin{center}
\includegraphics[width=1.0\hsize,angle=0]{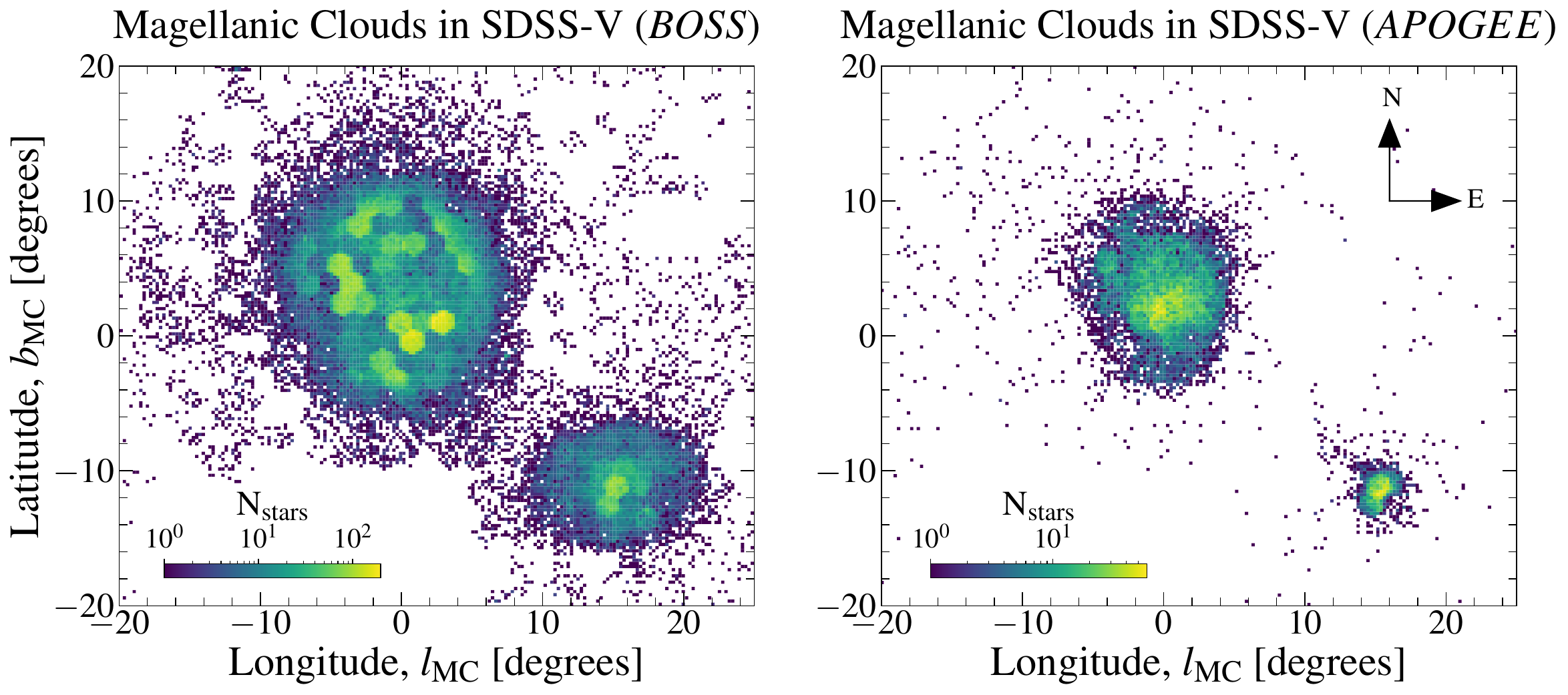}
\end{center}
\caption{SDSS-V's MGS mapping across the Magellanic cloud galaxies, showing the density of targets obtained with \textsl{BOSS} (left) and \textsl{APOGEE} (right) in an Magellanic cloud centered reference frame. In total, the MGS program will observe $\approx100,000$ red giant stars with \textsl{BOSS} and $\approx14,400$ asymptotic giant stars with \textsl{APOGEE}, and will cover the majority of the extent of both of these satellite galaxies.}
\label{fig:feh-sky}
\end{figure*}

Early spectroscopic efforts such as those by \citet{Olsen2011} and the MAgellanic Periphery Survey (MAPS; \citealt{Majewski2009}) provided important insights into the kinematics and extent of the Magellanic stellar populations, with the former finding a population of metal-poor stars in the inner LMC (possibly accreted from the SMC), and the latter discovering very extended stellar components of the MCs.
However, these surveys were either limited in spatial coverage or focused primarily on medium-resolution spectroscopy, providing little information about detailed element abundances. High-resolution spectroscopic studies have been especially sparse. For example, \citet{VanderSwaelmen2013} analyzed high-resolution spectra of only $\sim$100 LMC giants, representing the largest such dataset prior to the SDSS Apache Point Observatory Galactic Evolution Experiment \citep[APOGEE;][]{Majewski2017} observations.

The \textsl{APOGEE}-2S survey, as part of Sloan Digital Sky Survey-IV \citep[SDSS-IV;][]{Blanton2017}, began to address this gap by targeting approximately 5,000 red giant (RG) stars in the MCs with high-resolution near-infrared spectra (\citealt{Nidever2020}) using the second \textsl{APOGEE} spectrograph \citep{Wilson2019} on the Irenee du Pont telescope \citep{Bowen1973} at Las Campanas Observatory (LCO). These observations, reaching (S/N) greater than 100, have provided crucial information on element abundances and radial velocities out to radii of $\sim$9$^\circ$ in the LMC and $\sim$3$^\circ$ in the SMC \citep{Hasselquist2021,Povick2024,Povick2025a,Povick2025b}. However, \textsl{APOGEE}-2S’s deep observations of the MCs (9 hours per field in the LMC and 12 hours in the SMC) limited its spatial coverage. Only $\sim$33\% of the LMC’s main body was covered, and the spectroscopic footprint remained patchy, with very limited data beyond 10$^\circ$ from the LMC center (see \autoref{fig:apogee2map}).

To fully understand the internal structure and extended components of the MCs --- and to trace their dynamic interaction history --- a more comprehensive, spatially-complete spectroscopic survey is needed. The Sloan Digital Sky Survey-V \citep[SDSS-V;][]{Kollmeier2026} Magellanic Genesis Survey (MGS) was designed to meet this challenge. Taking advantage of the SDSS-V robotic Focal Plane System \citep[FPS;][]{Pogge2020}, which can reposition 500 robots in minutes, the program obtained high-resolution \textsl{APOGEE} spectra of approximately 14,400 oxygen-rich asymptotic giant branch (AGB-O) stars.
When combined with the SDSS-IV data, this amounts to high-resolution ($R\approx 22,500$) near-infrared \textsl{APOGEE} spectra of 22,580 giant stars in the inner MCs, offering contiguous coverage of their main stellar bodies \citep{jimenezarranz25a}. 
In parallel, the survey includes medium-resolution ($R$$\sim$2,000) optical \textsl{BOSS} spectroscopy \citep{Smee2013} of approximately 100,000 red giant (RG) stars across the MCs, selected using \gaia~DR3 \citep{GaiaDR3} proper motions and photometry.
\autoref{fig:feh-sky} shows the density of targets and extent covered by the MGS program, and Table~\ref{table:program} lists the details of the MGS.

\begin{figure*}[t]
\begin{center}
\includegraphics[width=1.0\hsize,angle=0]{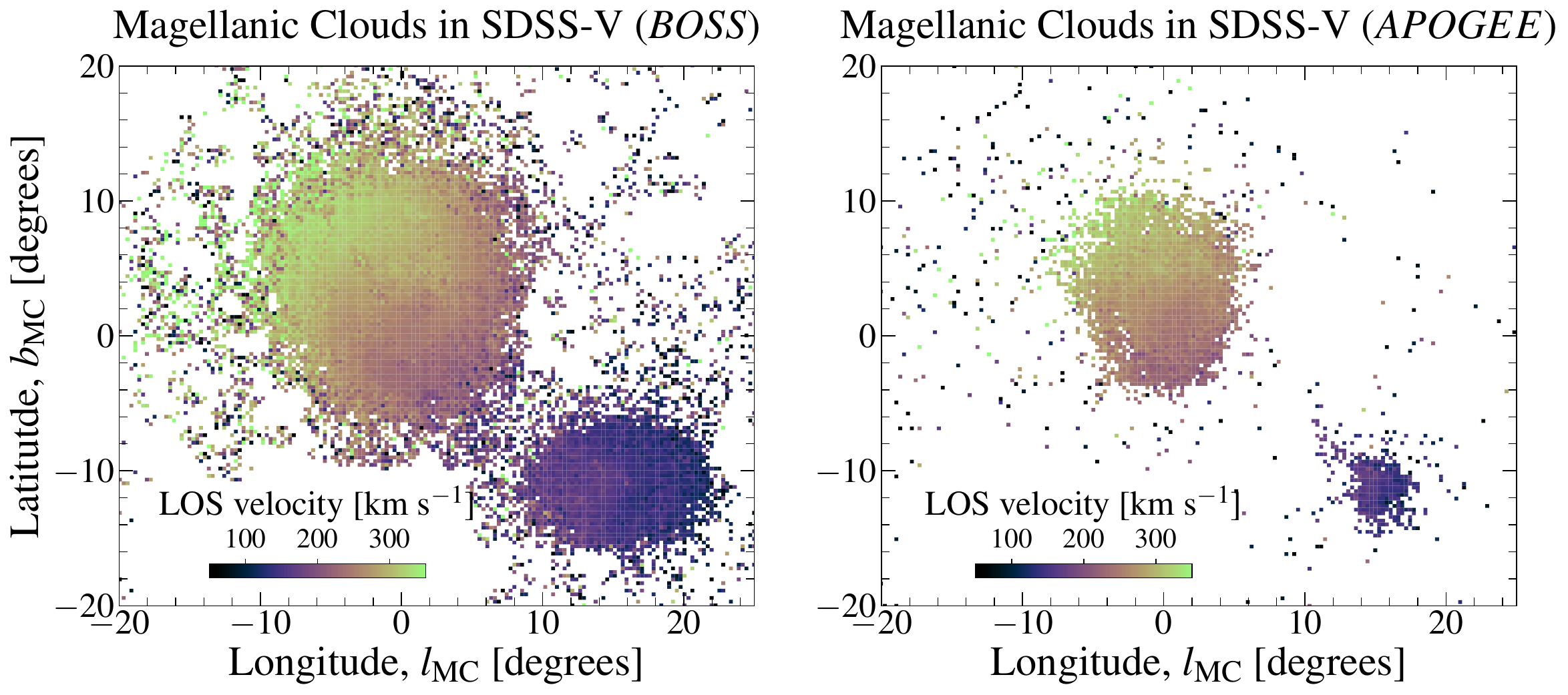}
\end{center}
\caption{Magellanic Genesis mapping of line-of-sight radial velocity across the Magellanic system. ({\em Left}) \textsl{BOSS} data; and ({\em Right}) \textsl{APOGEE} data.}
\label{fig:rv-sky}
\end{figure*}

\begin{deluxetable*}{lcccc}
\centering 
\tablecaption{SDSS-V Magellanic Genesis Survey}
\label{table:program}
\tablecolumns{5}
\tablewidth{500pt}
\tablehead{
  \colhead{Program} & \colhead{Instrument} & \colhead{Magnitude} & \colhead{N$_{\rm targets}$} \\
}
\startdata
AGB-O  &  \textsl{APOGEE}  &  10.0$<$$H$$<$12.9 & 14,400 \\
RGB     &  \textsl{BOSS}  &  14$<$$G$$<$17.5 & 100,000  \\
Evolved Massive Stars  &  \textsl{APOGEE} \& \textsl{BOSS}  &  8.5$<$$H$$<$13.0 & $\sim$300  \\
Symbiotic Binary Systems  &  \textsl{APOGEE}  &  10.0$<$$H$$<$13.0 & 7 
\enddata
\end{deluxetable*}

These data, combined with high-quality astrometry from \gaia~and spatially resolved star formation histories from deep optical photometry, will enable a transformative investigation of the chemo-dynamical structure of the MCs (see Figures \ref{fig:feh-sky} and \ref{fig:rv-sky}). In particular, the SDSS-V MGS will allow us to explore the extended stellar halos and substructures of the Clouds, measure chemical and kinematic gradients, identify signatures of past interactions, and ultimately constrain models of the Magellanic system’s dynamical evolution within the Local Group.

In this overview paper, we present the design, scope, and scientific objectives of the Magellanic Genesis Survey. 
In Section \ref{sec:strategy}, we describe the survey strategy and the target selection in Section \ref{sec:targets}. The final catalogs are described in Section \ref{sec:catalogs}, while initial results are presented in Section \ref{sec:results} and a brief summary is given in Section \ref{sec:summary}.


\section{The Magellanic Genesis Survey}
\label{sec:strategy}

We first discuss in more detail the strategy of the Magellanic Genesis survey and its three core components:
\begin{itemize}
    \item Mapping the kinematics and chemistry of the MCs main bodies with \textsl{APOGEE} observations of AGB stars.
    \item Mapping the kinematics and chemistry of the MCs peripheries with \textsl{BOSS} observations of RG stars.
    \item APOGEE+\textsl{BOSS} spectroscopy of rare evolved massive stars (supergiants, luminous blue variables, Wolf-Rayet, B[e]-stars), and symbiotic binary systems.
\end{itemize}


\subsection{Mapping the Kinematics and Chemistry of the Magellanic Main Bodies with APOGEE}
\label{subsec:apogee}

%
%


A central goal of the MGS is to construct a spatially contiguous, high-resolution spectroscopic map of the stellar populations across the MCs. Prior to SDSS-V, such coverage had not been fully achieved. The \textsl{APOGEE}-2S program in SDSS-IV \citep{Nidever2020} provided a major step forward by obtaining high-resolution $H$-band spectra of $\sim$5,000 RGB stars across the LMC and SMC. However, due to the use of ``pencil-beam'' fields, that survey ultimately covered only about one-third of the LMC’s main body, in a patchy and non-uniform distribution (see \autoref{fig:apogee2map}). Moreover, the outer regions and periphery of the Clouds were especially under-sampled.

To address these limitations, we designed the \textsl{APOGEE} component of the SDSS-V MGS to target a large, spatially uniform sample of asymptotic giant branch (AGB) stars. Specifically, we focused on the oxygen-rich AGB (AGB-O) stars, which are abundant, luminous in the near-infrared, and distributed throughout the entire extent of both the LMC and SMC. Their brightness allowed us to take advantage of the relatively short integration times available in SDSS-V ($\sim$15 min), enabling good signal-to-noise ($S/N$$\sim$$45$) observations for the majority of the sample without requiring multiple visits like for the fainter SDSS-IV RGB sample (which required 9--12 hours).

\begin{figure}[t]
\begin{center}
\includegraphics[width=1.03\hsize,angle=0]{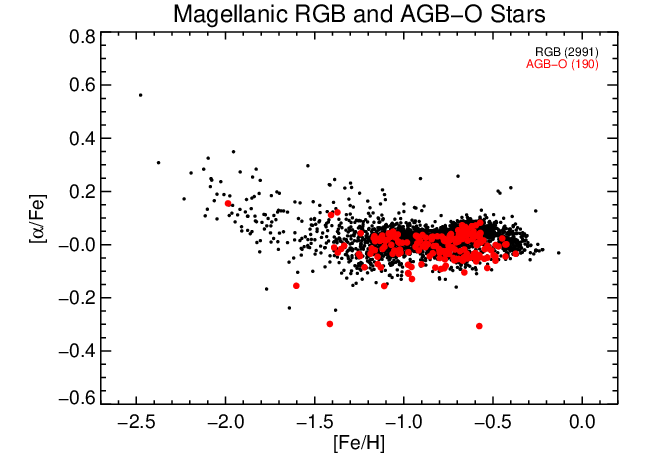}
\end{center}
\caption{The [$\alpha$/Fe] abundances for \textsl{APOGEE}-2S LMC and SMC RGB stars (black) and AGB-O stars (red), with S/N$\geq$80 and the ASPCAP \texttt{STARBAD} flag not set.
The chemical distribution of AGB-O stars matches that of the RGB stars quite well, though the former include few stars below [Fe/H]=$-$1.4.}
\label{fig:agbalphafe}
\end{figure}

Although AGB stars are in a late stage of stellar evolution (having undergone both first and second dredge-up), which makes interpreting their abundances more complicated, we verified using existing \textsl{APOGEE}-2S data of both RGB and AGB-O stars that they yield reliable stellar parameters, radial velocities, and detailed elemental abundances. \autoref{fig:agbalphafe} shows that the [$\alpha$/Fe] versus [Fe/H] distribution for those \textit{\textsl{APOGEE}-2S} AGB-O stars closely matches that of RGB stars in overlapping MC fields, particularly for metallicities [Fe/H] $\gtrsim -1.4$. 
Additional abundance comparisons across elements such as C, N, O, Mg, Al, Cr, Co, and Ni (see \autoref{fig:agbelemcomp} in the Appendix) demonstrate the consistency of AGB and RGB chemical signatures (although there are hints of small systematic differences), confirming that AGB-O stars are suitable tracers of the recent chemical enrichment history of the Clouds. However, AGB stars are predominantly produced by intermediate-age populations (0.1--6 Gyr) and are strongly biased against older ages \citep[$\gtrsim$6 Gyr;][]{Marigo2007}. Therefore, our AGB-O work will be focused on the recent evolution of the MCs.



While the AGB population naturally skews toward intermediate ages and is biased against very old and metal-poor stars, it samples the stellar populations formed over the last $\sim$6~Gyr, which corresponds to the epoch of peak star formation in both MCs. To extend our understanding to earlier epochs, we combined the new \textsl{APOGEE} AGB dataset with the RGB-rich \textsl{APOGEE}-2S sample. This synergy allows us to probe both the recent and ancient chemical evolution of the Clouds and to explore how these galaxies have evolved in response to their mutual interaction and their ongoing infall toward the MW.

\subsection{Mapping the Kinematics and Chemistry of the Magellanic Periphery with BOSS}
\label{subsec:boss}

A key objective of the MGS is to spectroscopically map the extended stellar envelopes and substructures of the MCs using optical spectroscopy of red giant stars (RGB and AGB). While the main bodies of the LMC and SMC have been increasingly well studied in photometric and spectroscopic surveys, the outskirts---which extend to $\sim$20\degr\ for the LMC and $\sim$12\degr\ for the SMC---remain poorly characterized, particularly in terms of stellar chemistry and kinematics.

\begin{figure*}[ht]
\begin{center}
\includegraphics[width=0.48\hsize,angle=0]{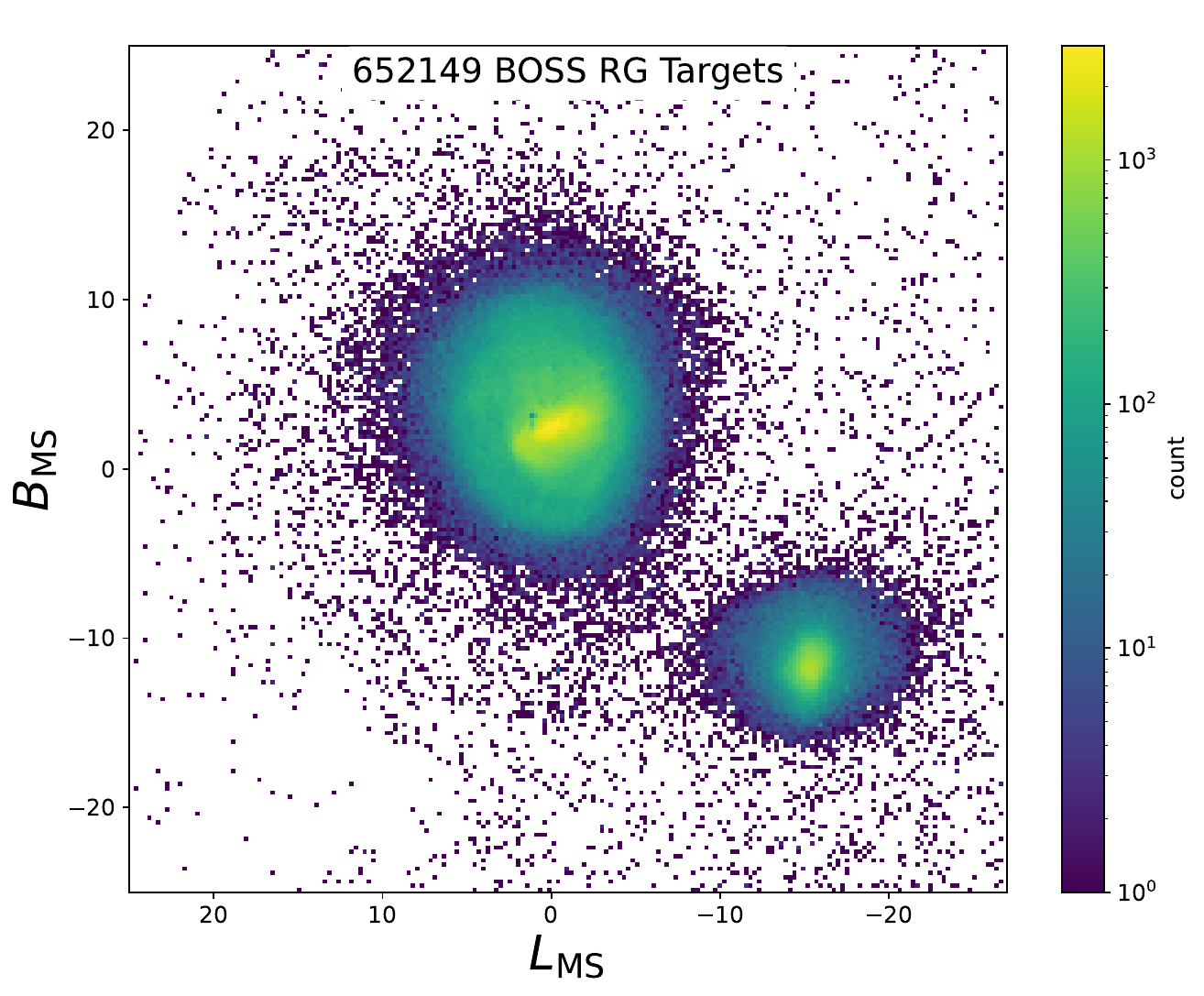}
\includegraphics[width=0.48\hsize,angle=0]{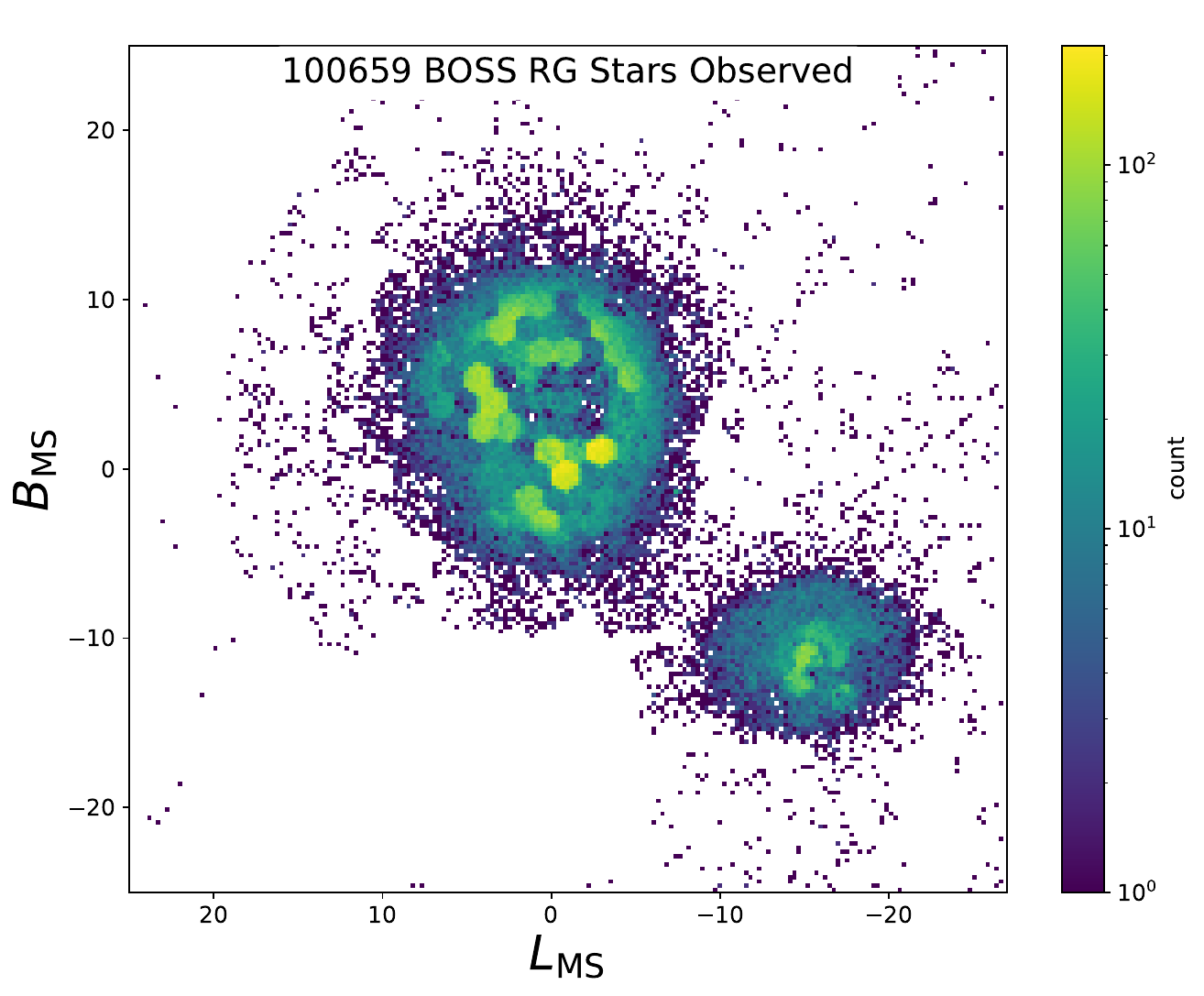}
\end{center}
\caption{Map of the MCs using RG stars selected with \gaia~DR3. ({\em Left}) Stars in the Magellanic periphery extend to at least $\sim$20\dgr around the LMC and organize into many diffuse substructures. The origin of these features is not well understood. ({\em Right}) Distribution of 100,659 MC RG stars observed by BOSS.}
\label{fig:bossmaps}
\end{figure*}

To address this gap, we obtained \textsl{BOSS} spectra for approximately 100,000 RG stars, with $\sim$13,000 in the outer periphery of the Clouds ($R_{\rm LMC}$~$\gtrsim$~8\dgr and $R_{\rm SMC}$~$\gtrsim$~4\degr). Targets were selected using \gaia~DR3 astrometry and photometry, with a magnitude limit of $G\lesssim17.5$, to isolate likely Magellanic giants. As shown in \autoref{fig:bossmaps} (left panel), the distribution of these stars reveals a wealth of complex, low-surface-brightness substructures, including extended spiral-like arms, potential tidal debris, and a diffuse stellar bridge connecting the LMC and SMC \citep{Belokurov2017, Nidever2024}. These features are suggestive of recent dynamical interactions and tidal stripping, but their origins and physical properties have remained elusive due to a lack of high-quality spectroscopic data.

The \textsl{BOSS} spectra, with resolution $R\sim2000$, provide radial velocities accurate to $\sim$5~km~s$^{-1}$ and allow the derivation of elemental abundances using spectral fitting tools such as \texttt{The Lux} \citep{Horta2025,Lux2025} or \texttt{The Payne} \citep{Ting2017}. For a subset of chemical elements (e.g., Fe, Mg, Ca, Si, Ni, Ti), abundance precisions better than 0.2~dex are achievable 
(see Figure 4 in \citealt{Ting2017}).
These data have enabled the construction of a contiguous map of kinematics and chemical composition across the outer Magellanic system, allowing us to trace gradients, identify coherent dynamical groups, and distinguish between native Magellanic populations and debris from tidal interactions.

In addition to the periphery, we also targeted a representative sample of RGB stars in the inner regions of both Clouds---out to radii of $R_{\rm LMC} \lesssim 8\degr$ and $R_{\rm SMC} \lesssim 4\degr$---as shown in \autoref{fig:feh-sky}. These inner samples serve as critical baselines for interpreting the outer populations. By anchoring the inner chemical and kinematical gradients, we can assess whether the outer substructures are consistent with tidal extension of the primary disks, accreted populations, or distinct evolutionary episodes.

This \textsl{BOSS} component of the Magellanic Genesis Survey significantly advances our understanding of the large-scale structure and evolutionary history of the MCs. It provides the first densely-sampled, chemically-tagged, and kinematically-anchored spectroscopic map of their outer stellar components, offering crucial insights into the interaction history between the Clouds themselves and with the MW halo.

\subsection{Evolved Massive Stars}
\label{subsec:massivestars}

The MGS includes a dedicated component focused on the rare and enigmatic class of evolved massive stars in the MCs, encompassing blue and yellow supergiants, luminous blue variables (LBVs), and stars transitioning into or out of the Wolf–Rayet (WR) phase. These objects represent key phases in the lives of the most massive stars, and they exhibit extreme phenomena including intense stellar winds, large-scale mass loss, spectral variability, and in the case of LBVs, dramatic outbursts and photometric brightening by several magnitudes.

\begin{figure*}[ht]
\begin{center}
\includegraphics[width=0.49\hsize,angle=0]{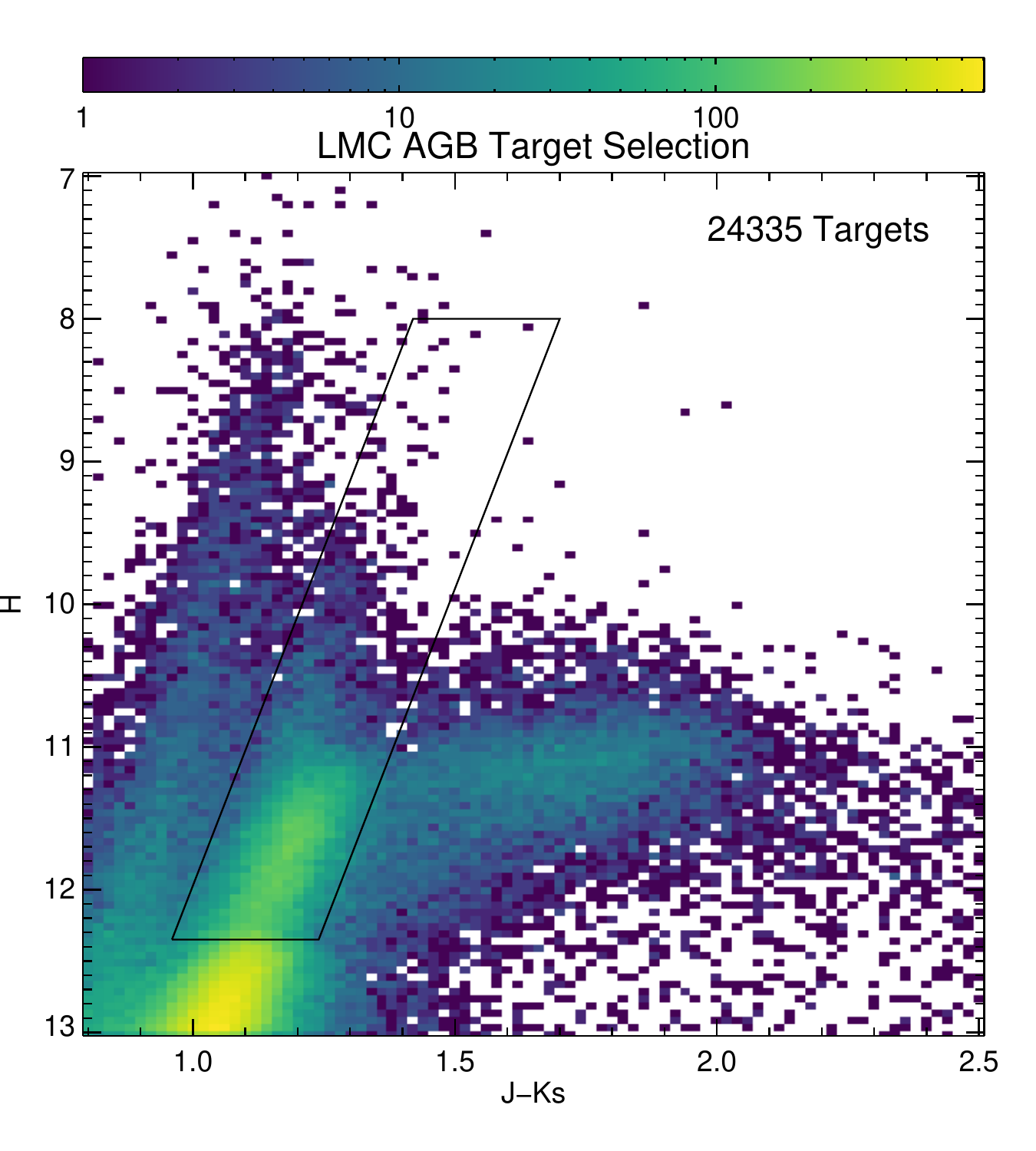}
\includegraphics[width=0.49\hsize,angle=0]{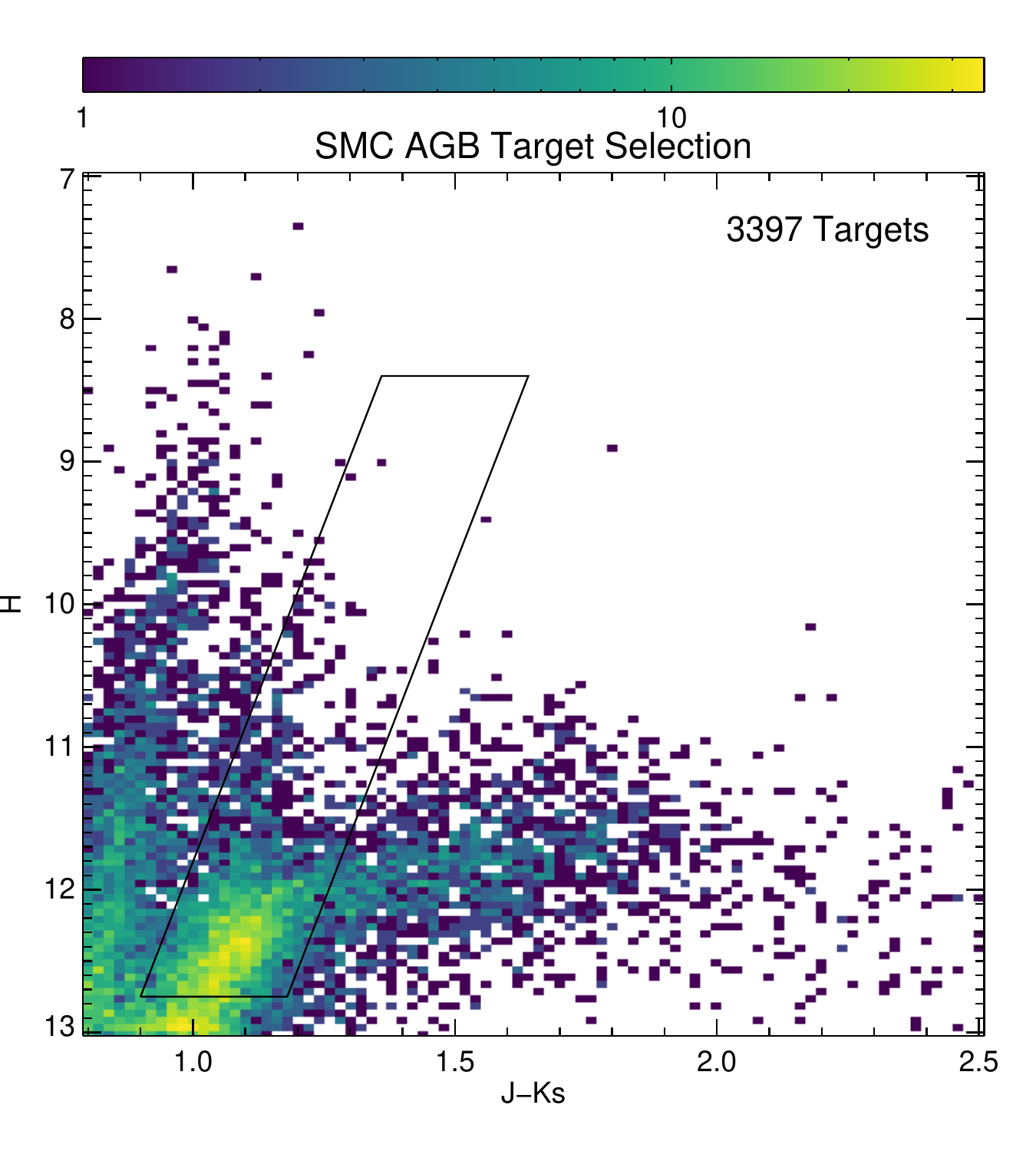}
\end{center}
\caption{Color magnitude diagram of 2MASS photometry in the LMC ({\em left}) and SMC ({\em right}) showing our AGB-O target selection strategy giving 24,335 targets in the LMC and 3,297 targets in the SMC.}
\label{fig:agbcmdselection}
\end{figure*}

The relative proximity of the LMC and SMC, combined with their rich massive star populations and relatively low foreground extinction, make them ideal environments for studying the late evolutionary stages of massive stars. These stellar populations span the full range of evolutionary phases and provide a unique laboratory for time-domain spectroscopic investigation.

In support of this science case, we obtained single-epoch \textsl{BOSS} and \textsl{APOGEE} spectra of hundreds of evolved massive stars (217 \textsl{APOGEE} and 160 BOSS, 60 with both), complementing the legacy dataset collected during SDSS-IV/\textsl{APOGEE}-2S.
As an example, the LBV RMC~71 in the LMC has been observed by \textsl{APOGEE} a total of 53 times and shows striking long-term variability in both photometry and spectra (see \autoref{fig:rmc71} in the Appendix). Despite just 12 \textsl{APOGEE} observations, the LMC LBV RMC 127 shows equally impressive variability (see Appendix \autoref{fig:rmc127} in the Appendix). Continued monitoring with SDSS-V has enabled further tracking of such behavior, expanding our ability to probe their evolutionary cycles.

The addition of \textsl{BOSS} spectroscopy provides crucial spectral typing capabilities in the blue optical region, allowing us to detect and interpret changes in line strengths and profiles associated with wind structure, disk formation, and eruptive events. When combined with the high-resolution near-infrared spectra from \textsl{APOGEE}, these data allow for a more comprehensive characterization of the physical conditions, element abundances, and dynamical processes in these stars.

Moreover, the synergy between \textsl{APOGEE}'s high spectral resolution and radial velocity precision---enhanced by the Fabry-Perot calibration system in SDSS-V---and the broad wavelength coverage of \textsl{BOSS} facilitates cross-calibration of elemental abundances between the near-infrared and optical regimes. This improves our understanding of the systematic differences between diagnostics and enriches our ability to compare results across different stellar populations and surveys.

This component of the MGS extends the time-domain baseline established by SDSS-IV and sets the stage for future synoptic spectroscopic studies of massive stellar evolution across the Local Group.

\subsection{Symbiotic Stars}
\label{subsec:symbiotic}

Another rare class of star system for which we can contribute significantly to constraining physical models is that of symbiotic binaries, which consist of a white dwarf accreting matter from a giant (RGB or AGB) star.  \textsl{APOGEE}-2 spectra of symbiotic binaries in MW satellites have already been used to make substantial strides in constraining and understanding wind-driven Roche Lobe Overflow in the Draco C1 system \citep{Lewis2020}, and an additional three known and five candidate symbiotic star systems (SySts) in the Clouds have \textsl{APOGEE}-2S spectra.  
For those having multi-epoch \textsl{APOGEE}-2S spectra, orbit fitting to the large amplitude radial velocity (RV) variability constrains the (typically several year period) orbits, a key step to establishing the white dwarf mass and binary separation, and, in turn, placing limits on the accretion mechanism and rate \citep{Washington2021}. 
SySts in \citet{Lewis2020} and \citet{Washington2021} are also found to exhibit line variability in their \textsl{APOGEE}-2 spectra (see \autoref{fig:lin358} in the Appendix), which correlates with orbital phase --- a discovery that points to yet another window of opportunity for constraining the SySt accretion physics by extending the \textsl{APOGEE}-2S time series database to cover the full periods of the (apparently elliptical) orbits.  

The goal of the new \textsl{APOGEE} observations of the eight confirmed systems and 16 others is to provide enough data to perform full orbital fitting of the binary, or, at a minimum, provide important constraints on the binary velocity semi-amplitude and spectral line variability of these systems over long temporal baselines. 




\begin{deluxetable*}{lcccc}
\centering 
\tablecaption{MGS \textsl{APOGEE} AGB-O Target Selection}
\label{table:agbtargetselection}
\tablecolumns{5}
\tablewidth{500pt}
\tablehead{
  \colhead{Type} & \colhead{Selection}
}
\startdata
Parallax ($\varpi$ [mas]) & $\varpi \leq$ 0.0 OR $(\varpi+0.025)/\sigma_\varpi \leq$ 5.0 \\
Proper Motion ($\mu$ [mas yr$^{-1}$]) & $\sqrt{(\mu_{L_{\rm MS}} - 1.8)^2 + (\mu_{B_{\rm MS}} - 0.4)^2}$ $<$ 1.2 \\
CMD (LMC) &  $J-K_{\rm s}$ = [0.96, 1.24, 1.70, 1.42], $H$ = [12.35, 12.35, 8.0, 8.0] \\
CMD (SMC) & $J-K_{\rm s}$ = [0.90, 1.18, 1.64, 1.36], $H$ = [12.75, 12.75, 8.4, 8.4]
\enddata
\end{deluxetable*}

\section{Target Selection and Exposure Times}
\label{sec:targets}

\subsection{\textsl{APOGEE} AGB-O Stars}
\label{subsec:targetsagb}


The MGS AGB-O stars were targeted using a CMD selection strategy similar to that used by the \textsl{APOGEE}-2S MCs survey (see Figure 3 of \citealt{Nidever2020}).
First, a spatial cut is used to select stars in the Magellanic region from the 2MASS PSC catalog \citep{Skrutskie2006},
r $< 30$\dgr from ($\alpha$, $\delta$) = (80.8925\degr, $-$72.1849\degr).
%
%
We then cross-match with the \gaia~DR3 catalog (closest match within 1\arcsec) and impose parallax and proper motion (in Magellanic Stream coordinates; \citealt{Nidever2008}) cuts given in Table~\ref{table:program}.


\autoref{fig:agbcmdselection} shows the 2MASS CMD of the LMC (left) and SMC (right) stars after \gaia~DR3 parallax and proper motion cuts have been applied. The black polygons show our AGB-O CMD target selection box (see Table~\ref{table:program}).  The selection box is shifted by 0.4 mag fainter for the SMC.
A line is used to spatially separate SMC from LMC stars:
$B_{\rm MS} \leq (-1.5728745 \times L_{\rm MS} -19.292601)$



\begin{figure*}[t!]
\begin{center}
\includegraphics[width=0.49\hsize,angle=0]{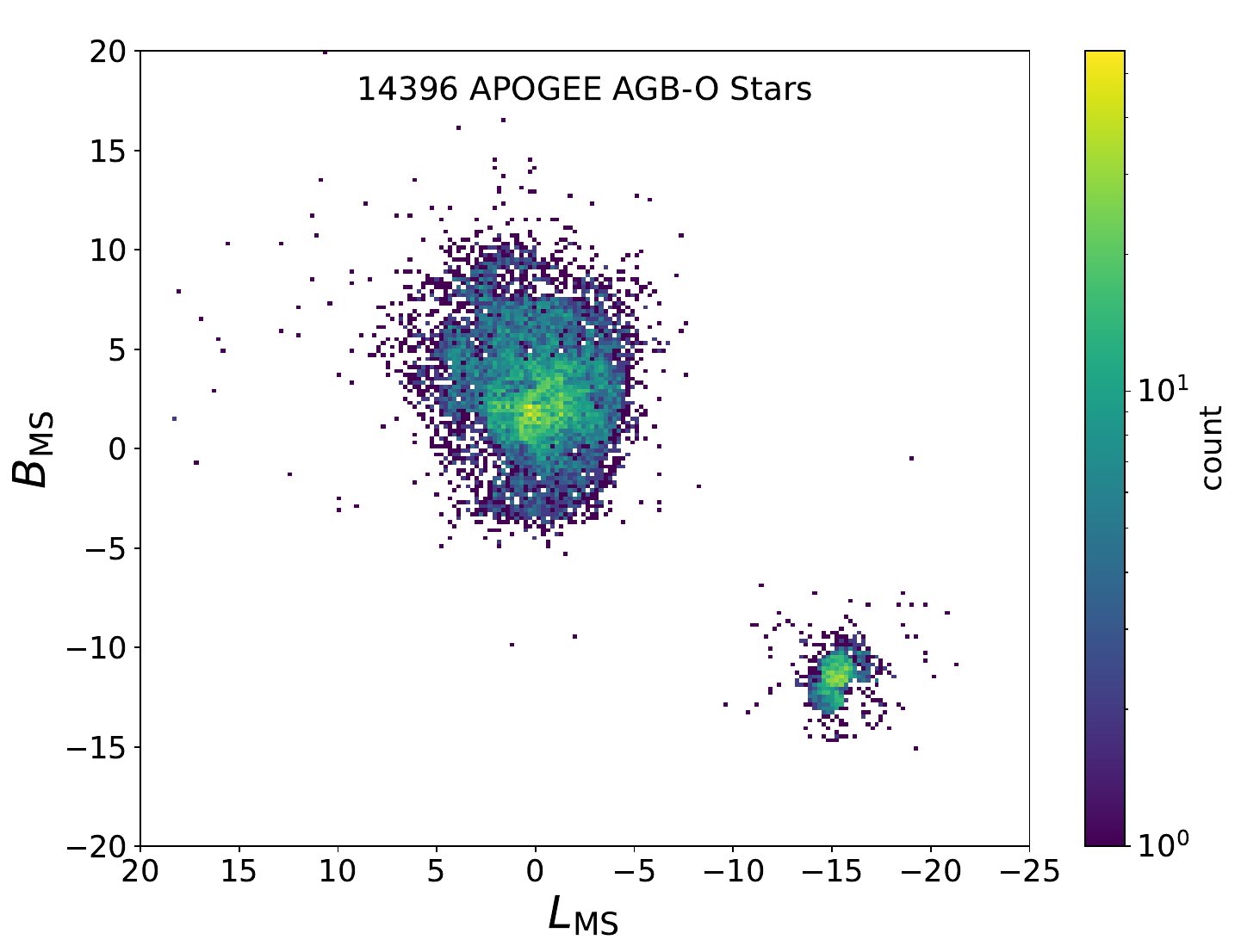}
\includegraphics[width=0.49\hsize,angle=0]{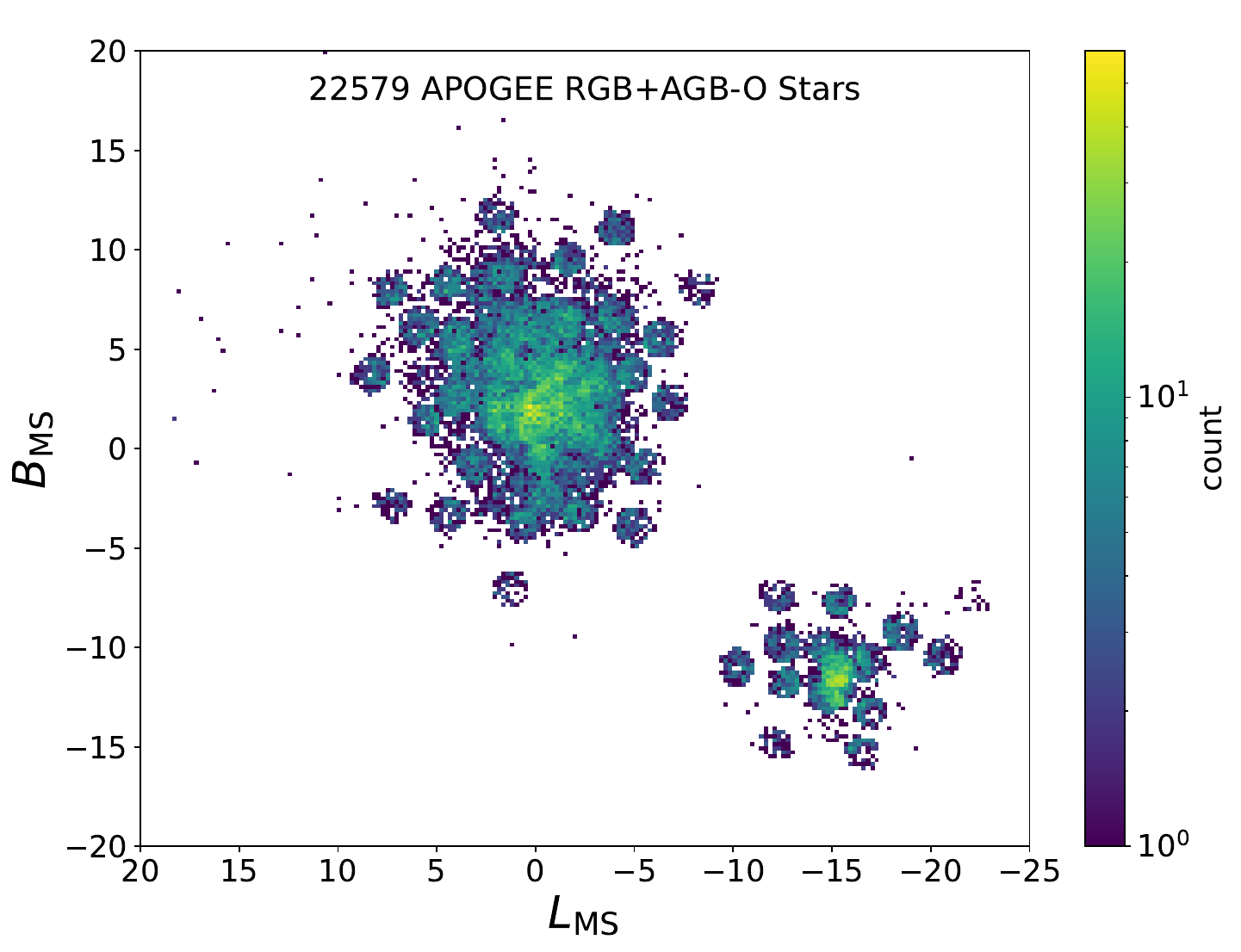}
\end{center}
\caption{Density map of Magellanic Genesis \textsl{APOGEE} AGB-O stars in Magellanic Stream coordinates \citep{Nidever2008}.  ({\em Left}) SDSS-V AGB-O stars (14,396), and ({\rm Right}) SDSS-IV and SDSS-V AGB-O and RGB stars (22,579).
}
\label{fig:agbmap}
\end{figure*}

These cuts select 28,156 targets, with 24,722 LMC targets and 3,434 SMC targets. This includes $\sim$2,000 AGB stars that were already observed in the \textsl{APOGEE}-2S MCs program that we use to cross-check our calibration against high-S/N spectra and will also allow us to study long-term variability.
\autoref{fig:agbmap} shows the full spatial distribution of the 28,156 AGB-O stars.
There is Python software to generate the target selection from public catalogs.\footnote{\url{https://github.com/sdss/target_selection/blob/main/python/target_selection/cartons/mwm_magcloud_agb.py}}

We benefited from experience gained by the SDSS-V Galactic Genesis program, which showed that precise elemental abundances can be derived for cool stars using \textsl{APOGEE} spectra with $S/N$$\sim$$45$. For our AGB sample, spectra with this quality yield abundances for all key \textsl{APOGEE} elements, while spectra with lower $S/N$ ($\sim$20--30) still allow us to measure stellar parameters, [Fe/H], mean $\alpha$-element abundances, and accurate radial velocities. These data are invaluable for dynamical modeling and the study of element abundance gradients.

\begin{deluxetable*}{lcccc}
\centering 
\tablecaption{MGS \textsl{BOSS} RG Target Selection}
\label{table:rgbtargetselection}
\tablecolumns{5}
\tablewidth{500pt}
\tablehead{
  \colhead{Type} & \colhead{Selection}
}
\startdata
Parallax ($\varpi$ [mas]) & $\varpi \leq$ 0.0 OR $(\varpi+0.025)/\sigma_\varpi \leq$ 5.0 \\
Proper Motion ($\mu$ [mas yr$^{-1}$]) & $\sqrt{(\mu_{L_{\rm MS}} - 1.8)^2 + (\mu_{B_{\rm MS}} - 0.4)^2}$ $<$ 1.2 \\
CMD (color) &  $(G_{BR}-G_{RP})_0$ = [0.95,1.2649,1.62255,1.9889,2.4164,3.1492,4.283,4.30,3.3411] \\
CMD (magnitude) & $G_0$ = [17.5,15.8064,15.2065,14.6065,14.0212,14.6358,15.6162,16.24,17.5]
\enddata
\end{deluxetable*}

\begin{figure*}[ht]
\begin{center}
\includegraphics[width=0.45\hsize,angle=0]{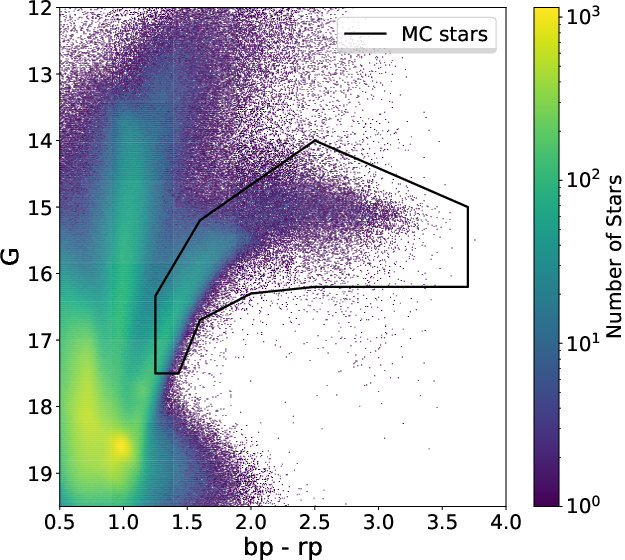}
\includegraphics[width=0.48\hsize,angle=0]{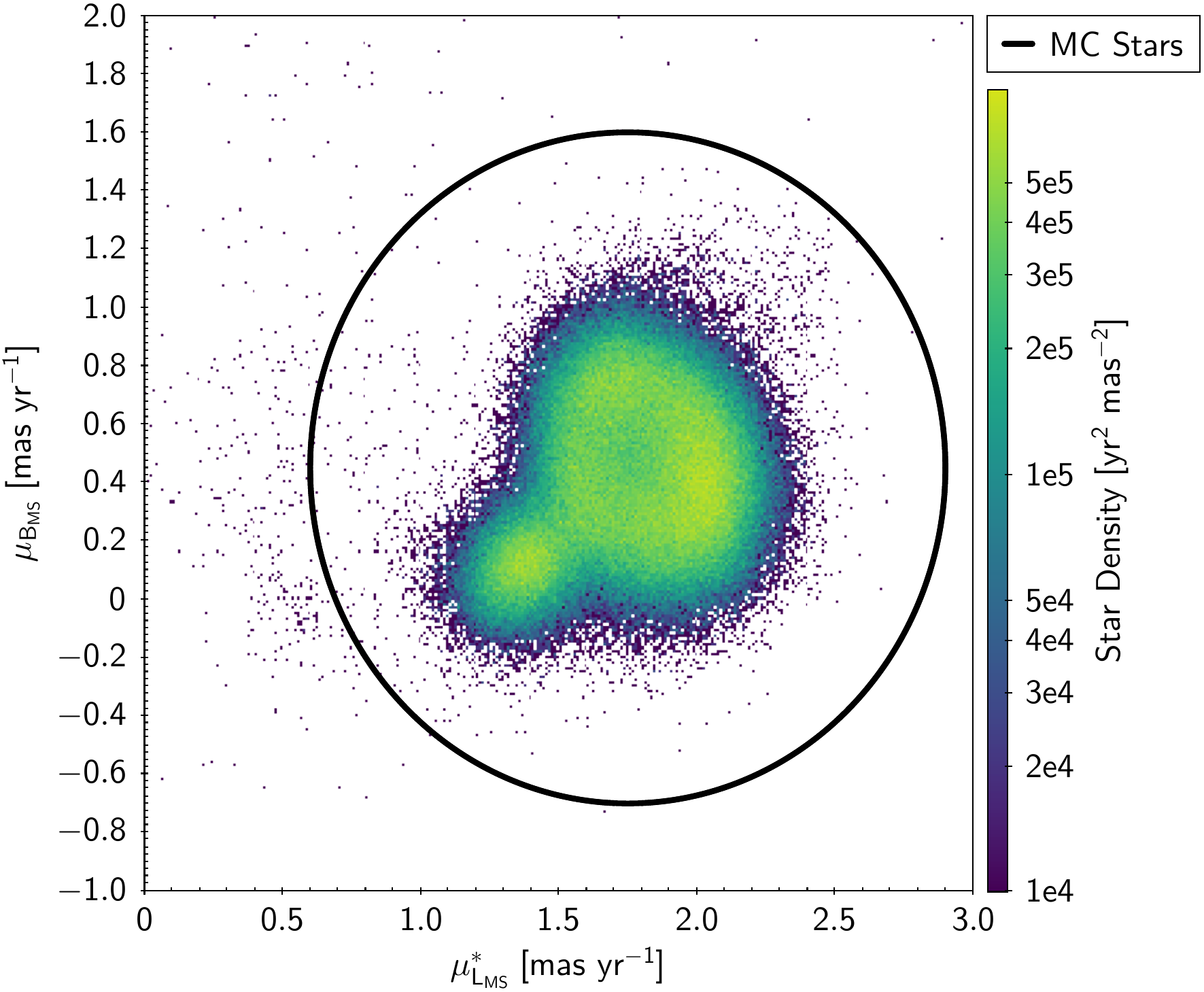}
\end{center}
\caption{\textsl{BOSS} red giant target selection with \gaia~DR3 data. ({\em Left}) Color magnitude diagram, and ({\em Right}) proper motion (in Magellanic Stream coordinates) target selection cuts.}
\label{fig:gaiacuts}
\end{figure*}

\subsection{\textsl{BOSS} RGB Stars}
\label{subsec:targetsrgb}

\autoref{fig:gaiacuts} shows a \gaia~DR3 color magnitude diagram (left panel) along with our cuts to select RGB and AGB stars (red giant stars; RG) to $G\approx17.5$. First, a spatial cut is used to select stars in the Magellanic region,
r $< 30$\dgr from ($\alpha$, $\delta$) = (80.8925\degr, $-$72.1849\degr).
After the spatial selection, we apply a CMD cut as in \cite{Belokurov2019} to obtain the stars inside the black polygon in \autoref{fig:gaiacuts} (left panel).
The photometry was dereddened using the rescaled SFD $E(B-V)$ \citep{Schlegel1998,Schlafly2011}.
We found that a decent ad hoc correction to the SFD $E(B-V)$ values in high extinction regions is to rescale them by 0.07$\times$  if $E(B-V)$ $\geq$ 0.1 mag for $R_{\rm LMC}$ $\leq$ 4.5\dgr and $R_{\rm SMC}$ $\leq$ 2.5\degr.



The proper motion selection can be seen in the right panel of \autoref{fig:gaiacuts} and is essential in the Magellanic periphery where the target density is low. The $\sim$2$\times$ improvement in proper motion uncertainties in \gaia~DR3 over DR2 improves the ability to select reliable Magellanic member stars.




The target selection criteria are summarized in Table~\ref{table:rgbtargetselection}. As mentioned above, there is Python software to generate the target selection from public catalogs.\footnote{\url{https://github.com/sdss/target_selection/blob/main/python/target_selection/cartons/mwm_magcloud_rgb.py}}

To achieve our goal of useful abundances ($\sigma_{\rm [X/Fe]}$ $<$ 0.2 dex) from these medium-resolution data, our goal was to obtain $S/N=10$ \textsl{BOSS} spectra which we achieve for $G=17.5$ stars in a single 15 minute visit.

\subsection{Evolved Massive Stars}
\label{subsec:targetsmassive}

Targeting for these stars used existing catalogs where spectral identification (and other classification criteria) had been predetermined (e.g., \citealt{Richardson2018,
Neugent2010, Neugent2012, Neugent2018, Zickgraf2006}. The final list has 1001 stars with $H<$13 which are generally concentrated in the central region of the MCs where current star formation is ongoing.

The new SDSS-V / \textsl{APOGEE} data in combination with the \textsl{APOGEE}-2 legacy dataset enables time domain science that can capture significant transitional evolutionary states that these stars undergo, though on unpredictable time scales.
The \textsl{BOSS} spectra provide new information that will be used for initial analysis and input into the SDSS-V modeling pipeline and provide a first \textsl{BOSS} spectrum for future continued observations.





\section{Final Catalogs}
\label{sec:catalogs}

All the data contained in these catalogs will be made publicly available and accessible via the \url{https://www.sdss.org/} website. As the \textsl{SDSS-V} survey is still taking observations, and these are yet to be analyzed, we describe here the timeline for the different data products. All optical spectra and derived stellar parameters and element abundance ratios for \textit{BOSS} targets will be available in the next public data release (namely, DR20), scheduled for July 2026. Conversely, near-infrared spectra and derived stellar parameters and element abundance ratios for \textit{APOGEE} stars will be made available in the final data release of the \textsl{SDSS-V} survey (DR21), scheduled for mid-2027. As was done with previous data releases, each catalog will contain a data model describing in detail its contents. However, it is intended that all \textit{APOGEE} catalogs and spectra closely resemble those of previous data releases, subject to current developments; conversely, the contents of the final \textit{BOSS} catalogs are still being decided. 

\begin{figure}[t]
\includegraphics[width=0.99\hsize,angle=0]{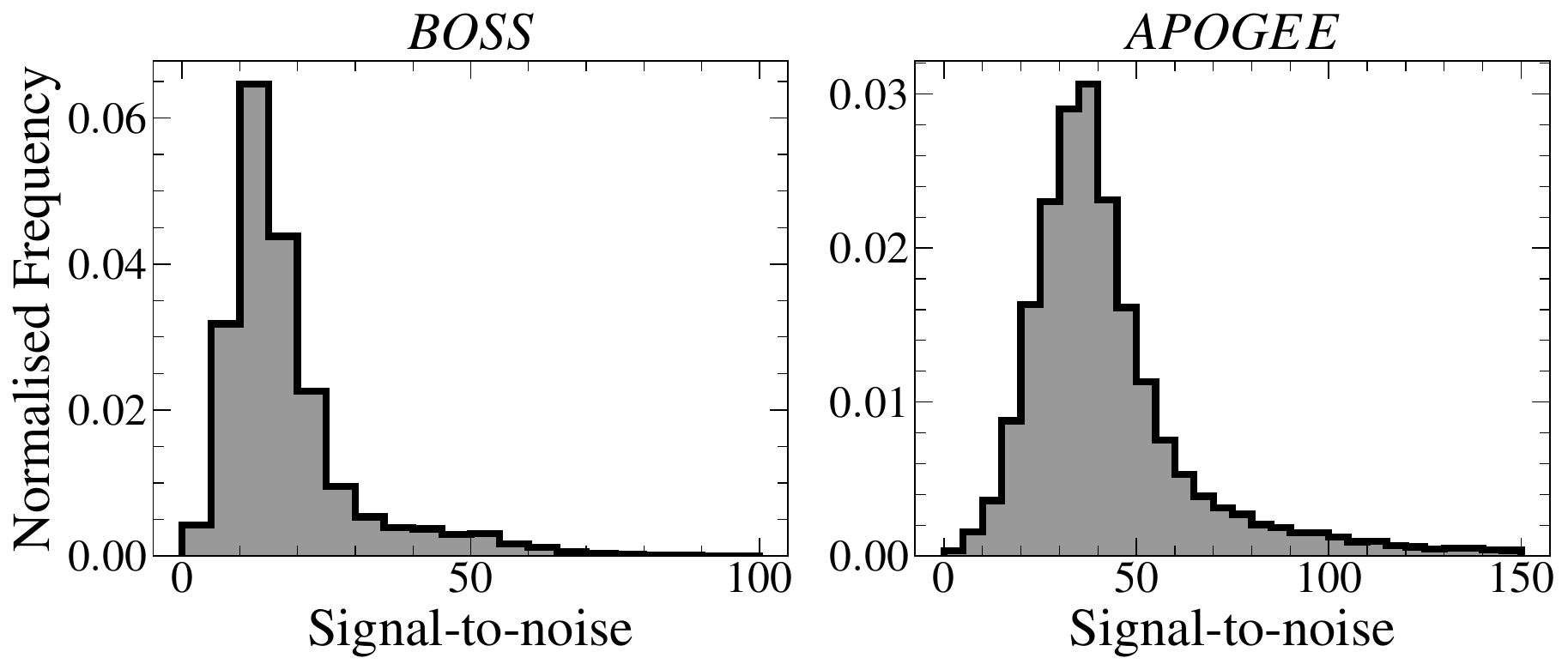}
\caption{S/N ratios for the ({\em left}) \textsl{BOSS} and ({\em right}) \textsl{APOGEE} stars.}
\label{fig:snr}
\end{figure}

The average spectral signal-to-noise ratio we obtain using both instruments is highlighted in Figure~\ref{fig:snr}; for \textit{BOSS}, we attain an average value of S/N~$\approx15$, and for \textit{APOGEE} S/N~$\approx45$. The average uncertainty attained in line-of-sight (LOS) velocities for the majority of the \textit{BOSS} targets is $\sim$5 km s$^{-1}$, and for \textit{APOGEE} targets is $\sim$0.06 km s$^{-1}$ (see \autoref{fig:rvsnr}).

In the following subsections, we describe the final individual MGS catalogs in more detail. 

\subsection{\textsl{APOGEE} Catalogs}
\label{subsec:apogeecatalogs}

Over the course of the MGS survey, we obtained new spectra for 14,396 AGB-O stars spanning the central and intermediate regions of the LMC and SMC.
The right panel of \autoref{fig:snr} shows the distribution of S/N of the AGB-O \textsl{APOGEE} spectra with a peak around S/N$\sim$47, and the RV uncertainty distribution is shown in the corresponding panel of \autoref{fig:rvsnr}.
In addition, \autoref{fig:agbmap} illustrates the broad spatial distribution of these targets, which collectively provide the first contiguous, high-resolution spectroscopic map of the inner MCs.

Together, the combined \textsl{APOGEE} AGB and RGB (22,579 stars) samples from SDSS-IV and SDSS-V constitute the most comprehensive high-resolution spectroscopic dataset ever assembled for the MCs. This dataset enables a wide range of science, including the construction of spatially resolved abundance maps, the study of chemo-dynamical substructures, and detailed modeling of galactic evolution in a dwarf-dwarf interaction context.

\subsection{\textsl{BOSS} Catalogs}
\label{subsec:bosscatalogs}

\begin{figure}[t]
\includegraphics[width=0.49\hsize,angle=0]{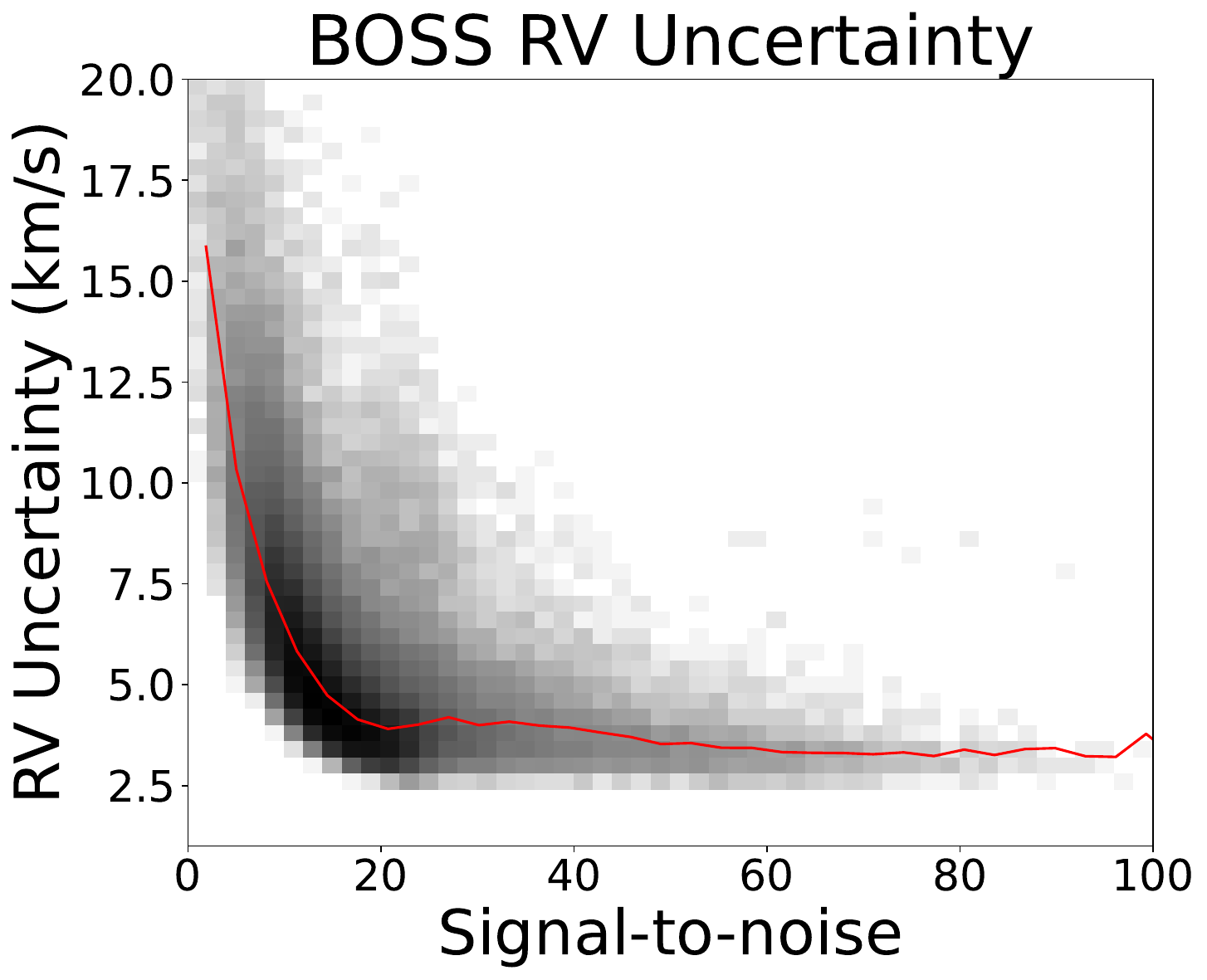}
\includegraphics[width=0.49\hsize,angle=0]{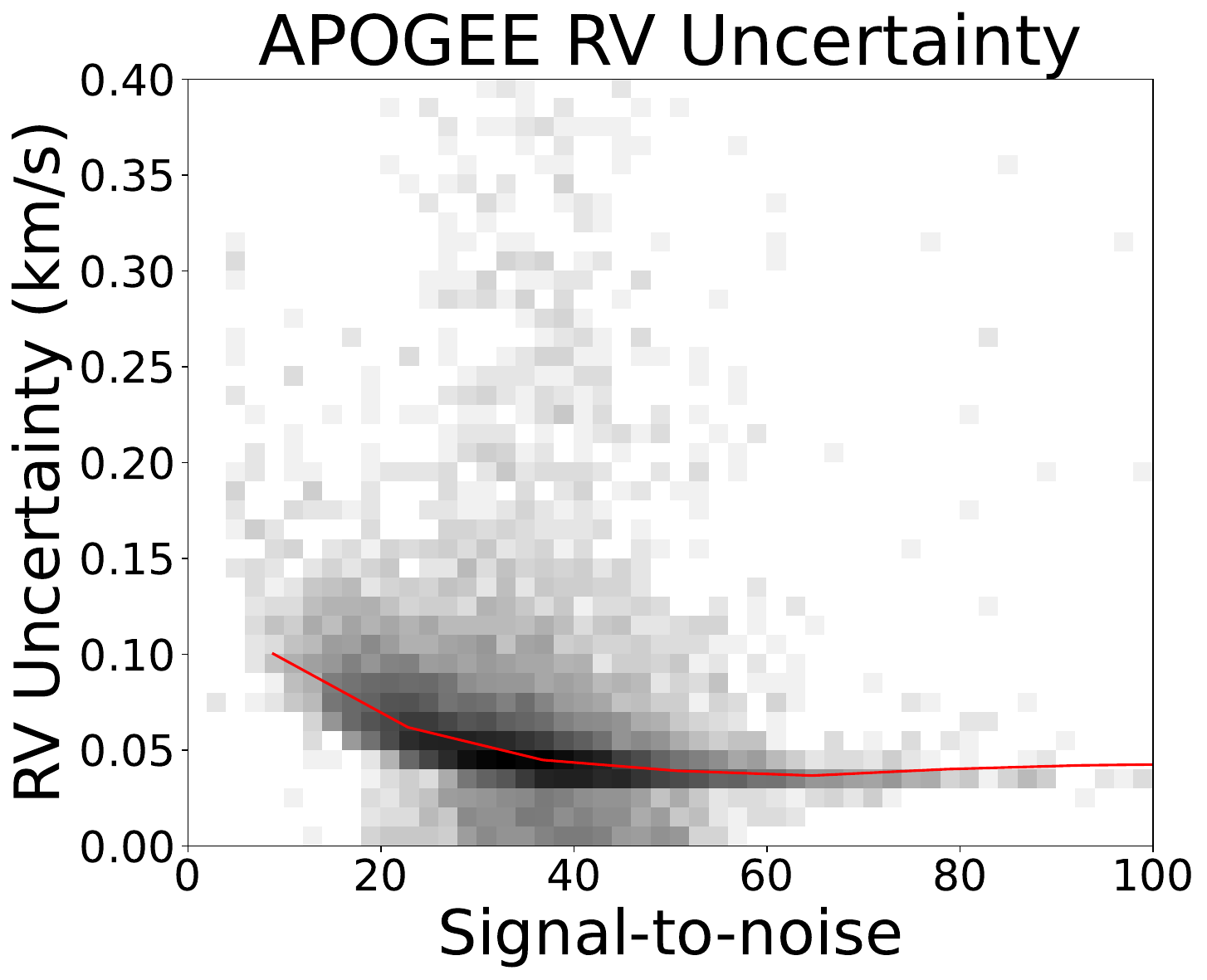}
\caption{RV uncertainty vs.~S/N ratios for the ({\em left}) \textsl{BOSS} and ({\em right}) \textsl{APOGEE} stars.}
\label{fig:rvsnr}
\end{figure}

We obtained \textsl{BOSS} spectra for 100{,}659 Magellanic RG stars. 
The left panel of \autoref{fig:snr} shows the S/N distribution of the \textsl{BOSS} RG spectra with a peak around $S/N$$\sim$22, and the RV uncertainty distribution is shown in the corresponding panel of \autoref{fig:rvsnr}.
Their distribution, as well as the distribution of all \textsl{BOSS} RG targets, is shown in \autoref{fig:bossmaps}. 
Comparing the \textit{BOSS} to \textit{APOGEE} RVs for the 2,554 stars in common, we find that there is a $+$7.31 \kms offset (\textit{BOSS}$-$\textit{APOGEE}) and scatter of 4.33 \kmse, indicating that the \textit{BOSS} RVs are accurate and precise.

\subsection{Massive and Symbiotic Star Catalogs}
\label{subsec:massivecatalogs}

Finally, \textsl{APOGEE} spectra have been obtained for 309 massive stars with 210 having \textsl{APOGEE} spectra, 159 \textsl{BOSS} spectra, and 60 stars having both.


\section{Initial Results}
\label{sec:results}

In this section, we present some initial results using the MGS dataset.

\subsection{Periphery Kinematic Maps}
\label{subsec:velocitymaps}

One of the key achievements of the MGS is its ability to map the kinematics of the MCs peripheries. Previous efforts, such as \citet{Olsen2011} and the MAPS survey \citep{Majewski2009}, were restricted to narrow, pencil-beam spectroscopic sight lines. In contrast, \textsl{BOSS} observations deliver contiguous RV coverage across the outskirts, enabling for the first time detailed kinematic mapping of the diffuse, low–surface-brightness structures surrounding the Clouds.

\begin{figure*}[ht]
\begin{center}
\includegraphics[width=0.95\hsize,angle=0]{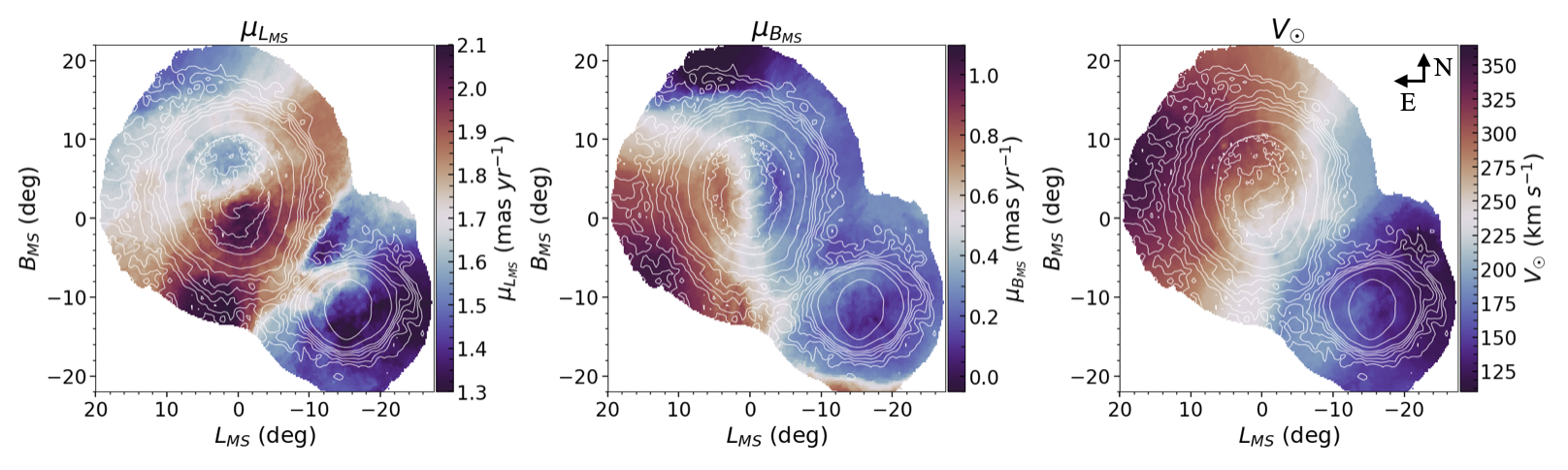}
\end{center}
\caption{Kinematics maps of red giant stars in the MCs extending to $\sim20^{\circ}$ from the LMC and $\sim12^{\circ}$ from the SMC. The PMs are from \gaia~DR3 and the \vhelio are from \textsl{BOSS} observations.}
\label{fig:velocitymaps}
\end{figure*}

The left two panels of \autoref{fig:velocitymaps} show the proper motion maps ($\mu_{L_{\mathrm{MS}}}$, $\mu_{B_{\mathrm{MS}}}$) from \textit{Gaia}~DR3 for the \textit{BOSS} targets, and the right panel displays the heliocentric RV ($V_{\odot}$) map from \textsl{BOSS} observations. The maps cover the MCs out to $\sim$20$^{\circ}$ from the LMC and $\sim$12$^{\circ}$ from the SMC, with spatial resolution ranging from $\sim$15{,}000 stars in the inner regions to $\sim$100 stars in the outer peripheries. They are generated using a nearest-neighbor search on a $0.3^{\circ}$ grid.

The stellar sample consists of high-probability Magellanic RGB stars, selected using a parallax cut of $\varpi < 0.20$~mas and a local 3$\sigma$ kinematic filter \citet{Oden2025}, which effectively removes most non-Magellanic contaminants. White contours trace the stellar number density from Oden et al., highlighting key morphological features such as the LMC bar, spiral arm, northern arm, southern hooks, and the SMC northern overdensity. The inner LMC bar/arm contours are derived following the method of \citet{JimenezArranz2023}. Cardinal directions are indicated in the upper-right panel for reference.

The inner LMC, centered at ($L_{\mathrm{MS}}$, $B_{\mathrm{MS}}$) $\approx$ (0.0, 3.0), displays pronounced rotational signatures in the left two panels. In contrast, the SMC, centered at ($L_{\mathrm{MS}}$, $B_{\mathrm{MS}}$) $\approx$ ($-$15.0, $-$13.0), shows weak or no evidence of ordered inner rotation—consistent with previous measurements \citep[e.g.,][]{vandermarel2014,Choi2022,JimenezArranz2023}. Beyond the main stellar disk of the LMC, the velocity field becomes increasingly complex, revealing coherent streams and asymmetries associated with low–surface-brightness features such as the northern arm and southern hooks. These features exhibit clear kinematic departures from simple disk rotation, pointing to tidal perturbations and ongoing dynamical interactions between the LMC, SMC, and the Milky Way. The kinematic maps presented here provide a crucial baseline for future dynamical modeling of the Clouds’ stellar peripheries and for reconstructing their recent interaction history.

\subsection{LMC Vertical Structure}
\label{subsec:velocitymaps}

Another main result of the Magellanic Genesis Survey is the study of the vertical structure and kinematics of the LMC using both \textsl{SDSS-IV/V} and \gaia~data \citep{jimenezarranz25a}. Previous investigations of the internal kinematics of the LMC had provided a detailed view of its structure, largely thanks to the exquisite proper motion data supplied by the \gaia~mission \citep[e.g.,][]{Schmidt2020,luri20,Niederhofer2021,Choi2022,Cullinane2022a,Niederhofer2022,JimenezArranz2023,jimenezarranz24a,Vijayasree2025}. However, RVs --- the third component of stellar motion --- were available for only a small subset of the existing \gaia~data, limiting studies of the kinematics perpendicular to the LMC disk plane.

In \citet{jimenezarranz25a}, new \textsl{SDSS-IV/V} RV measurements were combined with existing \gaia~DR3 data, increasing the 5D phase-space sample by almost a factor of three -- reaching up to 80,000 stars. Using this unprecedented dataset, the authors interpreted and modeled the vertical structure and kinematics across the LMC disk. The two main results of this work were: (1) the first identification of the supershell LMC 4 in the (vertical) kinematics space of resolved LMC stars, and (2) the production of 3D representations of the LMC disk shape, showing that it is not a flat plane in equilibrium, but that the central bar region is tilted relative to a warped outer disk.

\begin{figure*}[t!]
\centering
\includegraphics[width=1\textwidth]{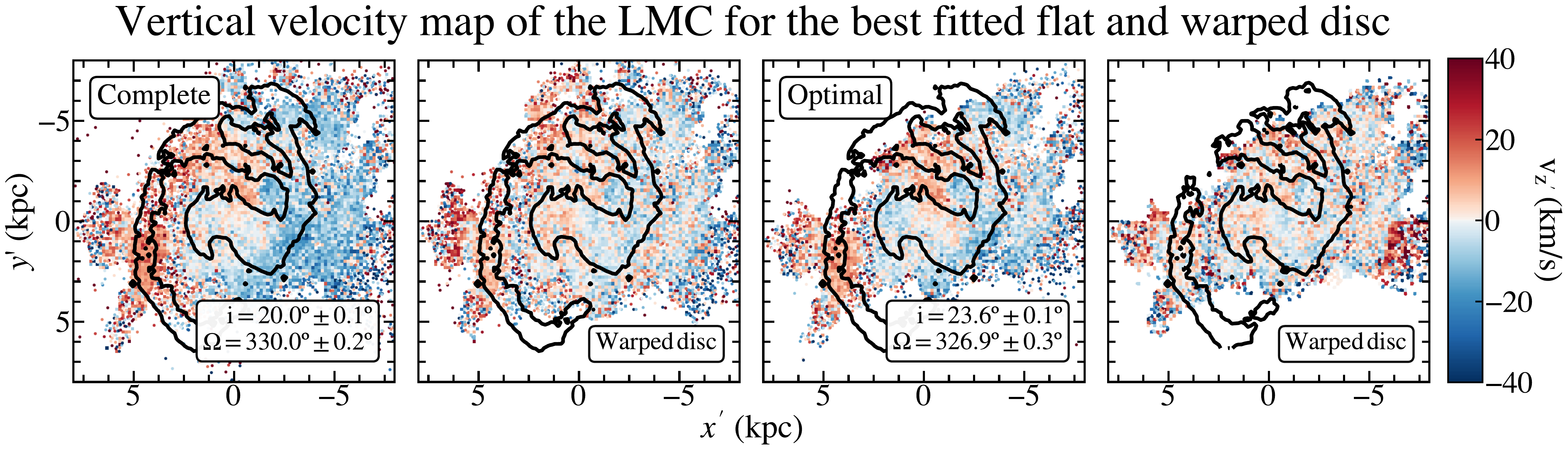}
\caption{Comparison of the median vertical velocity maps for the different LMC Combined samples (top: complete sample; bottom: optimal sample) obtained using SDSS-V and \textit{Gaia} data with different assumptions about the LMC disc plane, taken from \citet{jimenezarranz25a}. \textit{Left}: Results for the best-fitting flat plane, obtained by minimising the RMS vertical velocity $v_z'$, with viewing angles listed in the panels. \textit{Right}: Results for the best-fitting warped plane. When comparing the top and bottom panels, the reader should note that the overdensity lines differ slightly because
each LMC sample uses a different disc plane.}
\label{fig_OJA}
\end{figure*}

\autoref{fig_OJA} shows the resulting median vertical velocity maps for different SDSS-V-\textit{Gaia} samples in the LMC. The first and third panel show the best-fitting flat plane, obtained by minimising the RMS vertical velocity, $v_z'$, with viewing angles listed in each panel. The second and fourth panels show the results for the best-fitting warped plane. While the newly determined disc plane (by definition) provides a lower RMS vertical velocity than the one determined by \citet{luri20}, the median maps in the first and third panels continue to show significant residual structure. For further details, see \citet{jimenezarranz25a}.



\subsection{Abundance maps}
\label{subsec:abundmaps}

\begin{figure*}[t]
\begin{center}
\includegraphics[width=0.45\hsize,angle=0]{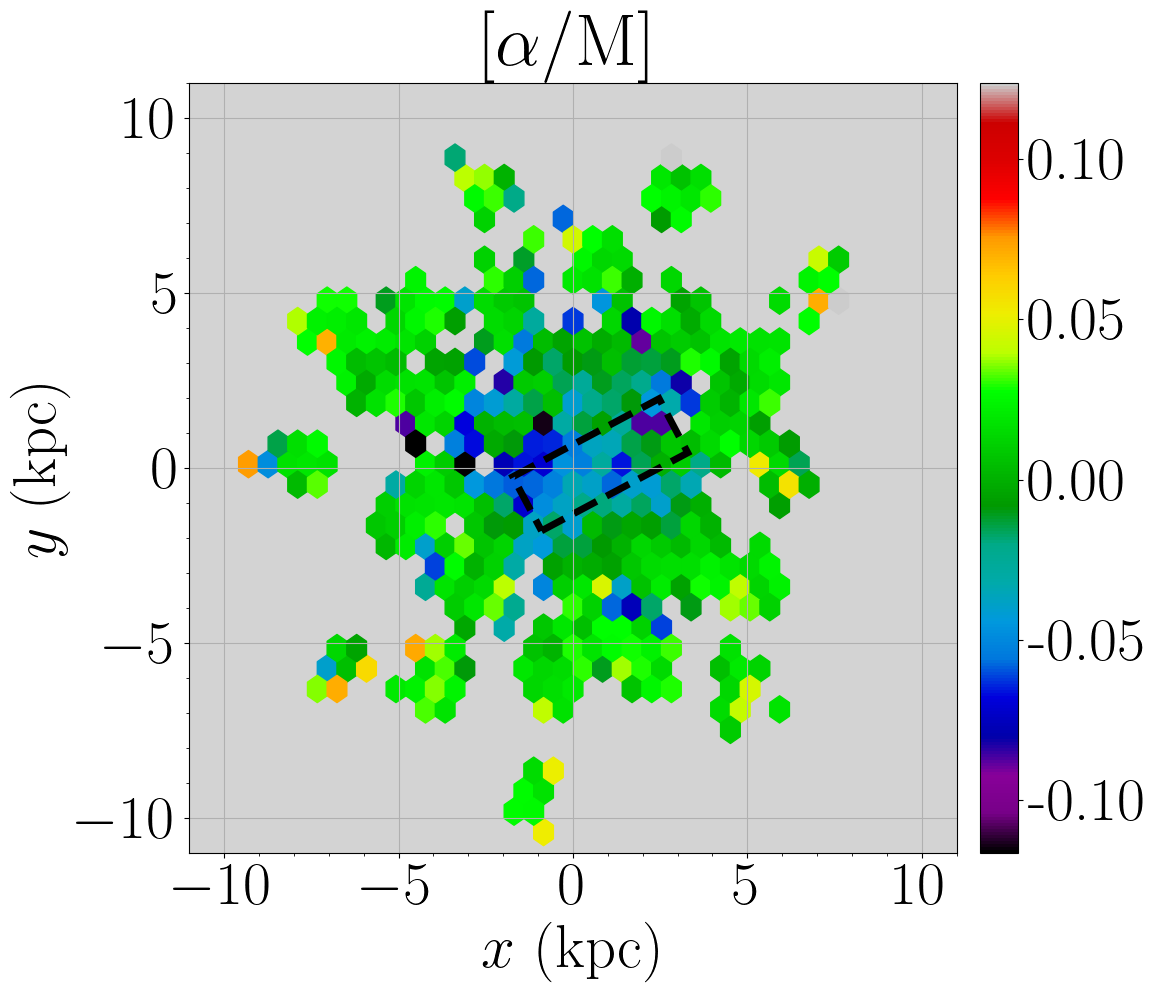}
\includegraphics[width=0.45\hsize,angle=0]{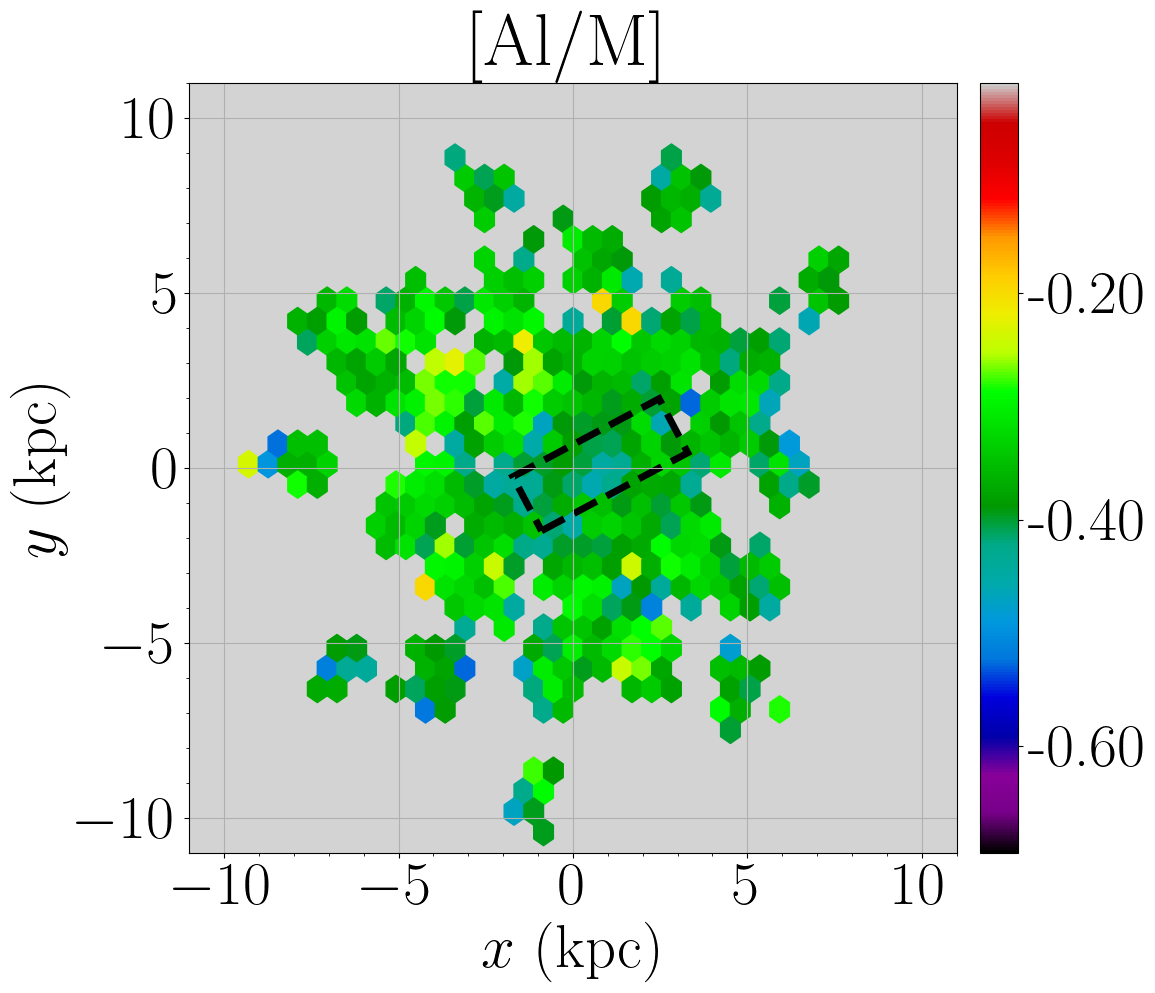} \\
\includegraphics[width=0.45\hsize,angle=0]{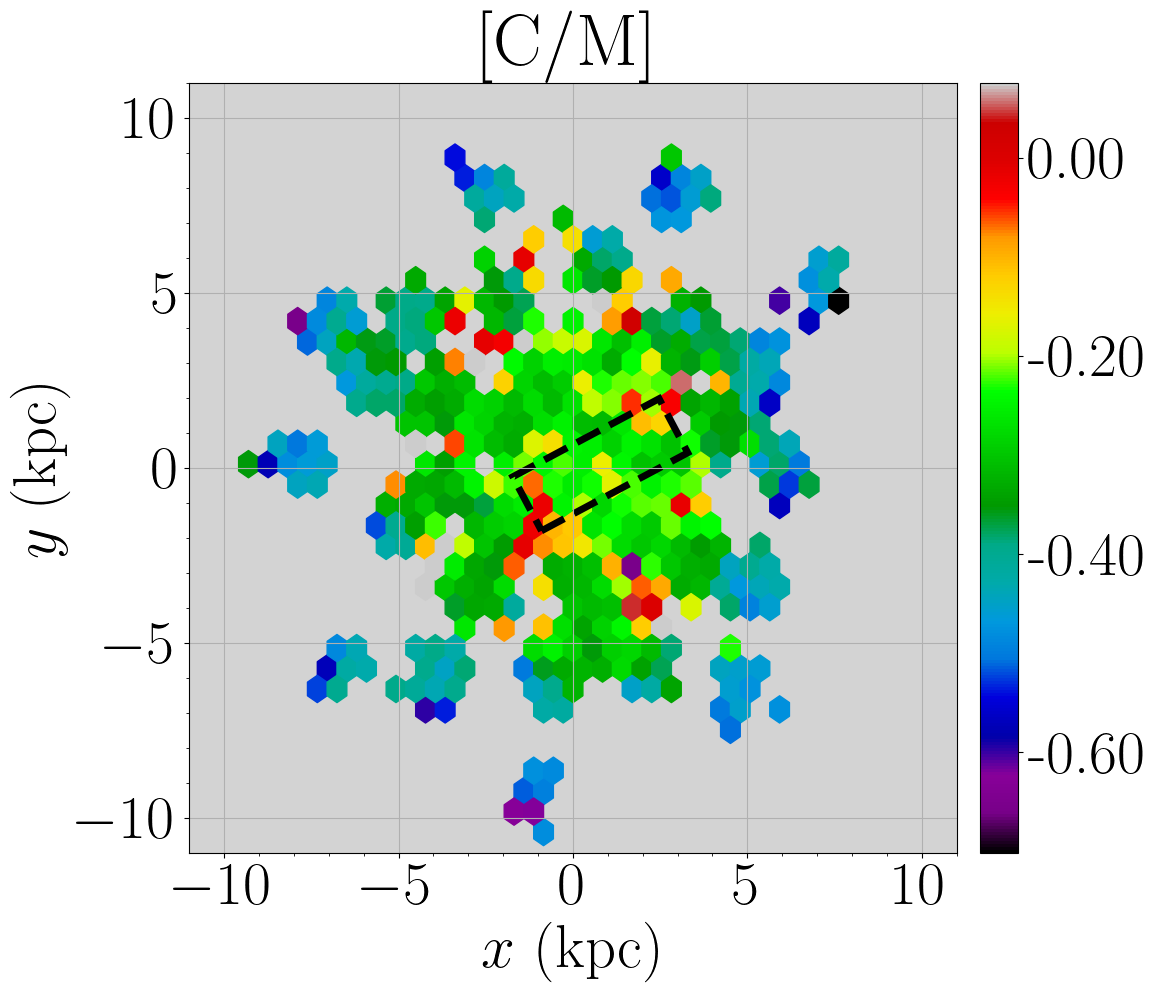}
\includegraphics[width=0.45\hsize,angle=0]{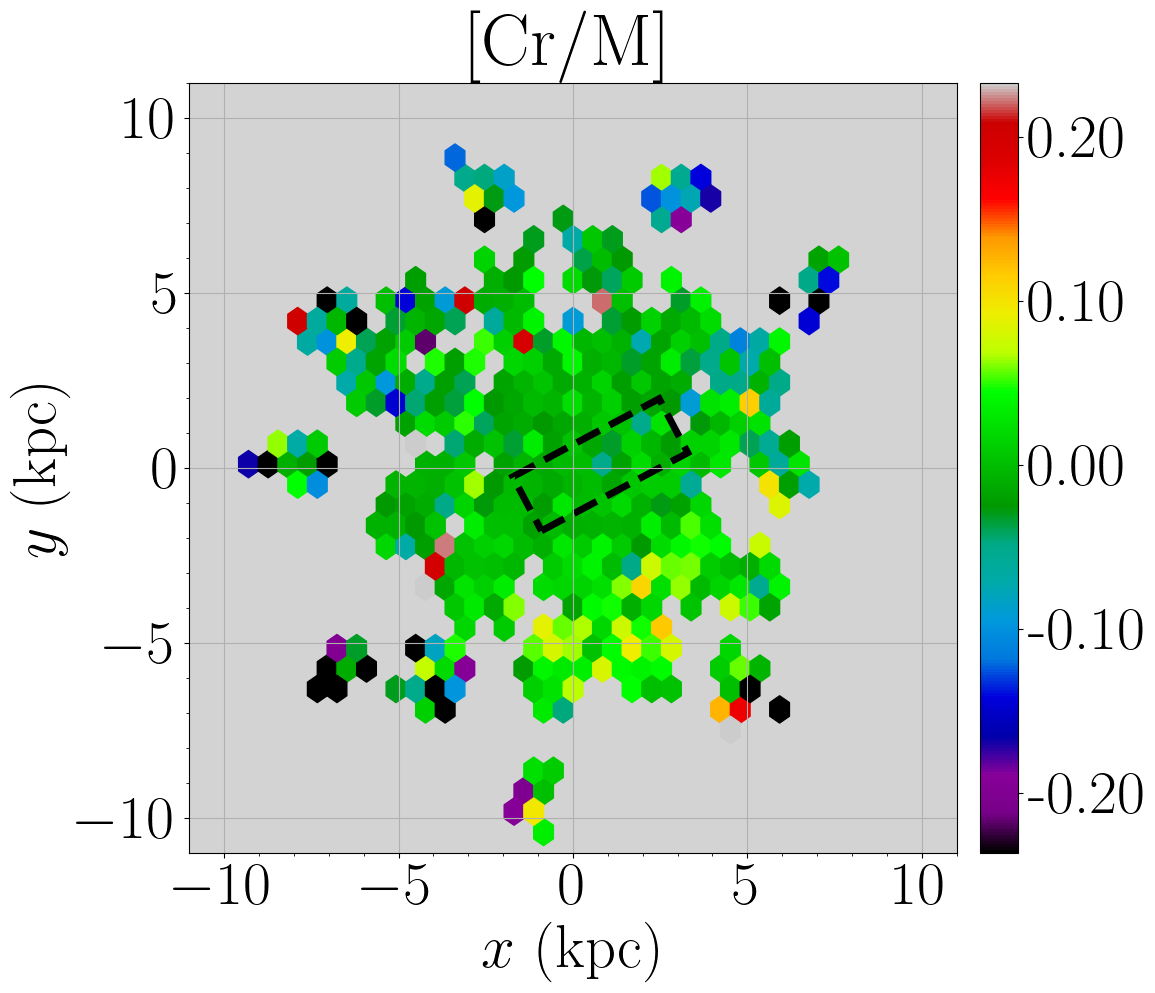} \\
\end{center}
\caption{MGS \textsl{APOGEE} abundance maps of the RGB and AGB stars for {[$\alpha$/M]}, {[Al/M]}, {[C/M]}, and {[Cr/M]}.  The dashed black rectangle approximately shows the location of the off-centered bar in the LMC.}
\label{fig:abundmaps}
\end{figure*}

By combining the spectroscopy of both \textsl{APOGEE} and \textsl{BOSS} with the addition of many new targets, the Magellanic Genesis survey allows for a more contiguous wide-field view of the MCs (see \autoref{fig:feh-sky}). The new expanded coverage fills in many gaps that were present in SDSS-IV, enabling a more detailed mapping of the spatial distribution of chemical elements in the two galaxies. Maps such as these can give deeper insights into the chemical evolution and the relative contributions of different nucleosynthetic pathways that inform the present day compositions of the MCs. Initial abundance maps for a few select abundance ratios ({[$\alpha$/M]}, {[Al/M]}, {[C/M]}, and {[Cr/M]}) can seen in \autoref{fig:abundmaps}.


The {[$\alpha$/M]} values represent the full $\alpha$-abundance of a star (i.e., a combination of O, Mg, Ca, and Ti) and is enriched into the ISM primarily through Type II SNe \citep{nomoto2013nucleosynthesis,weinberg2019chemical}. It is measured using a global fit of a stellar spectrum. In \autoref{fig:abundmaps}, the upper left panel shows the initial {[$\alpha$/M]} map from MGS, with a clear positive radial gradient. This could potentially be explained by low metallicity gas being funneled into the LMC from the SMC due to an interaction \citep[e.g.,][]{Bekki2007,Besla2012,zivick2019}. Additionally, the LMC bar and 30 Doradus stand out in the map. 

The next panel in the upper right of \autoref{fig:abundmaps} is {[Al/M]}, that in the \textit{APOGEE} spectra is measured from individual atomic lines. Aluminum is an odd-Z element produced in exploding massive stars, with its yields being metallicity dependent \citep[e.g.,][]{woosley1995massive,kobayashi2006chemical,nomoto2013nucleosynthesis}. This means that the {[Al/M]} will trace a similar pattern to $\alpha$-elements at low metallicity, but will become distinguishable at higher [M/H]. Overall, we find that there is no difference in the average [Al/M] across the LMC disk.


Much like {[$\alpha$/M]}, {[C/M]} is measured from a global fit of a stellar spectrum. Carbon is produced in both Type II SNe \citep{kobayashi2006chemical} and also intermediate mass AGB stars \citep{ventura2013agb}. It is also an important element for giants stars because it is sensitive to internal mixing. This means {[C/M]} is important on the scale of nucleosynthesis of a galaxy as well as individual stellar evolutionary processes. In the lower left panel of \autoref{fig:abundmaps}, we find that [C/M] appears to vary radially, with the inner regions of the LMC disk being more [C/M]-rich when compared to the outskirts. With the current data, this radial trend appears to be qualitatively equal across all azimuth angles.

In the last panel of \autoref{fig:abundmaps} is a map of the {[Cr/M]} for the LMC, which is determined from individual atomic lines in the \textsl{APOGEE} spectrum. Chromium is a member of the iron peak elements with an atomic number close to iron. It is mostly produced in Type Ia SNe, but also with metallicity dependent yields from Type II SNe \citep{kobayashi2006chemical}. Similarly to when inspecting [Al/M], we find no clear radial or azimuthal gradient in [Cr/M]. However, we do note that the southern LMC disk appears to have slightly more enhanced values of [Cr/M] on average when compared to the rest of its disk. However, these average differences are small, on the order of $\sim0.05$ dex.

\section{Summary}
\label{sec:summary}

The MGS, conducted as part of SDSS-V, has successfully executed a comprehensive, spectroscopic mapping of the Clouds using both the \textsl{APOGEE} and \textsl{BOSS} instruments. The survey achieved its primary goal of producing a spatially contiguous, high-resolution view of the stellar chemistry and kinematics across both the inner and outer regions of the Clouds.

Utilizing the luminous population of oxygen-rich asymptotic giant branch (AGB-O) stars, we obtained \textsl{APOGEE} $H$-band spectra for approximately $\sim$14,400 stars, effectively covering the entire main bodies of the LMC and SMC. These observations have enabled the construction of a detailed, two-dimensional chemo-kinematic map of the inner regions of the Clouds, revealing spatial variations in chemical abundance patterns and stellar motions at unprecedented resolution.

In parallel, the survey observed $\sim$100,000 red giant stars across the MCs (including the outer periphery) using \textsl{BOSS} optical spectroscopy. These data extend the chemical and kinematic mapping to radii of $\sim$20\degr\ for the LMC and $\sim$12\degr\ for the SMC, probing their extended stellar halos and the complex, diffuse substructures that have been recently discovered in wide-area photometric and astrometric surveys. These outer components are key to understanding the past interactions between the Clouds and their ongoing evolution within the MW halo.

In addition to the large population studies, the survey also obtained single-visit spectra of $\sim$300 evolved massive stars and eight symbiotic binary systems in the Clouds. These observations, conducted with both \textsl{APOGEE} and \textit{BOSS}, provide valuable snapshots of rare, late-stage stellar systems and contribute to the long-term monitoring of these objects, bridging the gap between the \textit{\textsl{APOGEE}-2} legacy sample and future time-domain spectroscopic programs.

Together, the MGS has produced the most complete spectroscopic dataset to date for the Magellanic system. When combined with \gaia~astrometry and existing photometric star formation histories, the dataset provides a powerful resource for reconstructing the formation history, chemical enrichment, and dynamical interactions of the Clouds. The results from this survey not only enhance our understanding of these nearby dwarf galaxies, but also serve as critical benchmarks for interpreting more distant systems where such detail is inaccessible. Specifically, the data delivered by the MGS program will be useful to tackle the following open questions:
\begin{itemize}
    \item What are the chemical enrichment histories, and by inference the star formation histories, of the LMC and SMC galaxies? and how do they compare to galaxies ranging in mass and formation timescales?
    \item What is the resolved chemical-kinematic structure of the LMC and SMC? Is it well ordered? How can we use it to infer their assembly history and past interactions? 
    \item What effect, and to which extent, do these twin galaxies have on the dynamical structure of the MW halo?
    \item How do the LMC and SMC compare to other satellite galaxies in the Local Group?
    \item How do massive stars evolve in their final stages of evolution? and how does this affect star formation in lower-mass galaxies?
\end{itemize}



\acknowledgements
D.L.N. acknowledges support from NSF grants AST 1908331 and 2408159. D. H. acknowledges support by the UKRI Science and Technology Facilities Council under project 101148371 as a Marie Sk\l{}odowska-Curie Research Fellowship.

Funding for the Sloan Digital Sky Survey V has been provided by the Alfred P. Sloan Foundation, the Heising-Simons Foundation, the National Science Foundation, and the Participating Institutions. SDSS acknowledges support and resources from the Center for High-Performance Computing at the University of Utah. SDSS telescopes are located at Apache Point Observatory, funded by the Astrophysical Research Consortium and operated by New Mexico State University, and at Las Campanas Observatory, operated by the Carnegie Institution for Science. The SDSS web site is \url{www.sdss.org}.

SDSS is managed by the Astrophysical Research Consortium for the Participating Institutions of the SDSS Collaboration, including the Carnegie Institution for Science, Chilean National Time Allocation Committee (CNTAC) ratified researchers, Caltech, the Gotham Participation Group, Harvard University, Heidelberg University, The Flatiron Institute, The Johns Hopkins University, L'Ecole polytechnique f\'{e}d\'{e}rale de Lausanne (EPFL), Leibniz-Institut f\"{u}r Astrophysik Potsdam (AIP), Max-Planck-Institut f\"{u}r Astronomie (MPIA Heidelberg), Max-Planck-Institut f\"{u}r Extraterrestrische Physik (MPE), Nanjing University, National Astronomical Observatories of China (NAOC), New Mexico State University, The Ohio State University, Pennsylvania State University, Smithsonian Astrophysical Observatory, Space Telescope Science Institute (STScI), the Stellar Astrophysics Participation Group, Universidad Nacional Aut\'{o}noma de M\'{e}xico, University of Arizona, University of Colorado Boulder, University of Illinois at Urbana-Champaign, University of Toronto, University of Utah, University of Virginia, Yale University, and Yunnan University.

This work has made use of data from the European Space Agency (ESA) mission
{\it Gaia} (\url{https://www.cosmos.esa.int/gaia}), processed by the {\it Gaia}
Data Processing and Analysis Consortium (DPAC,
\url{https://www.cosmos.esa.int/web/gaia/dpac/consortium}). Funding for the DPAC
has been provided by national institutions, in particular the institutions
participating in the {\it Gaia} Multilateral Agreement.

This publication makes use of data products from the Two Micron All
Sky Survey, which is a joint project of the University of Massachusetts
and the Infrared Processing and Analysis Center/California Institute of
Technology, funded by the National Aeronautics and Space Administration
and the National Science Foundation.

This research has made use of NASA’s Astrophysics Data System Bibliographic Services; the arXiv preprint server operated by Cornell University; and the SIMBAD databases hosted by the Strasbourg Astronomical Data Center.

Portions of the writing and editing in this paper were assisted with the help of the ChatGPT large language model (OpenAI, 2025). The tool was used to improve clarity, phrasing, grammar, and structure of draft text, while all scientific content, interpretations, and conclusions remain entirely the responsibility of the authors.

\bibliographystyle{apj}
\bibliography{ref_og.bib}

\newpage

\appendix


\begin{figure*}[ht]
\begin{center}
\includegraphics[width=0.45\hsize,angle=0]{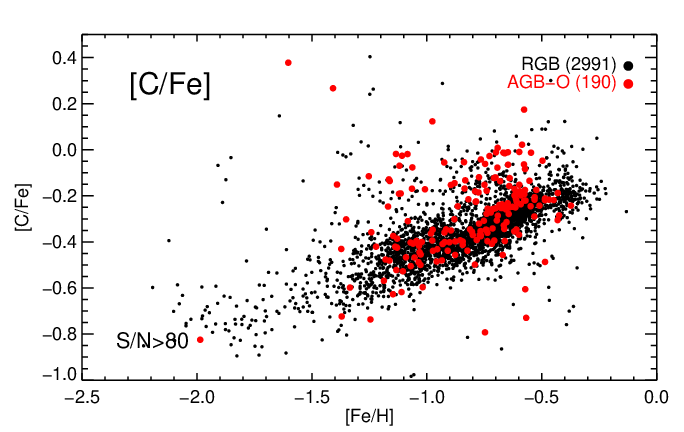}
\includegraphics[width=0.45\hsize,angle=0]{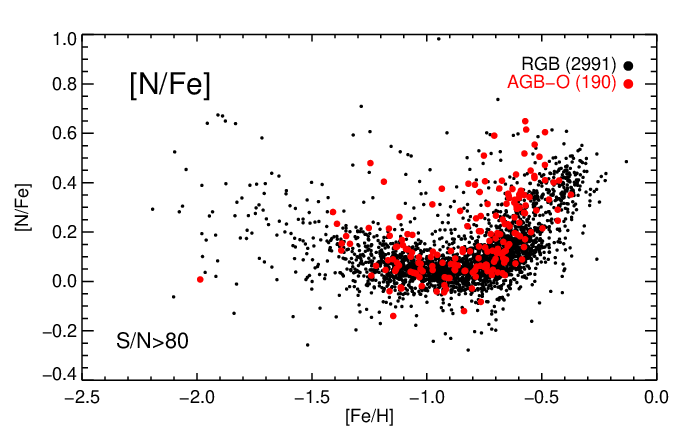} \\
\includegraphics[width=0.45\hsize,angle=0]{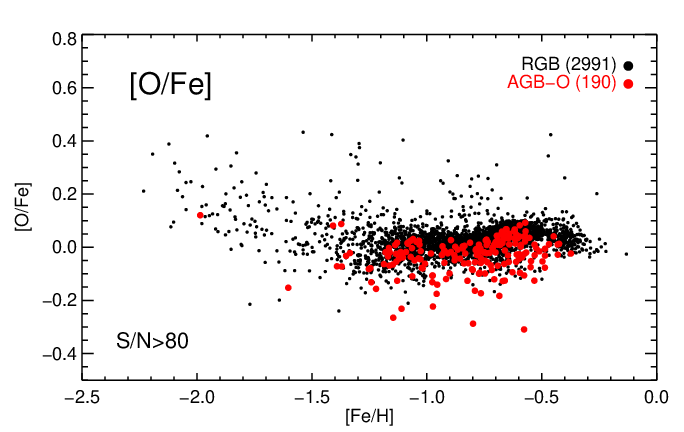}
\includegraphics[width=0.45\hsize,angle=0]{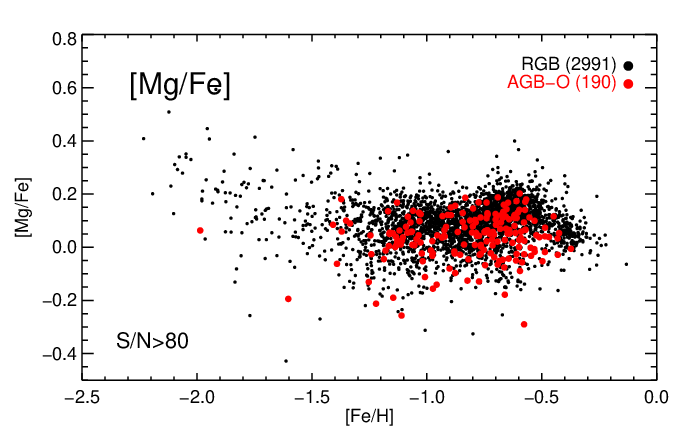} \\
\includegraphics[width=0.45\hsize,angle=0]{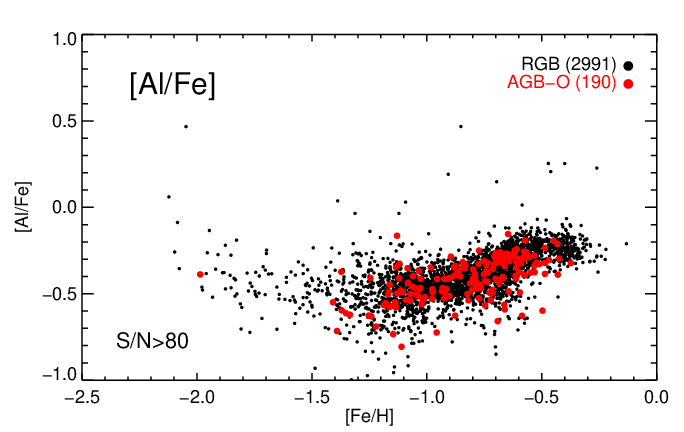}
\includegraphics[width=0.45\hsize,angle=0]{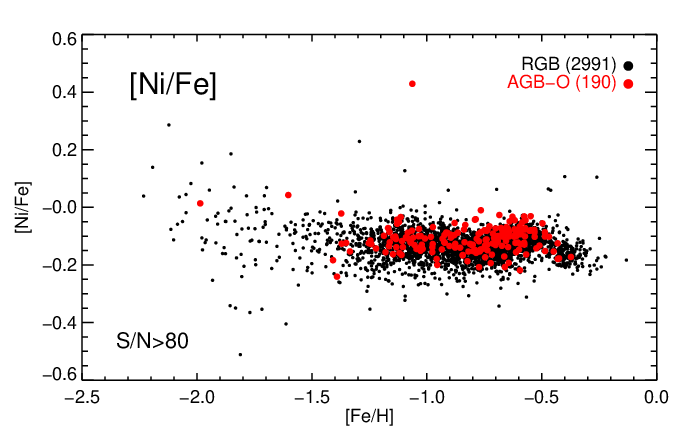}
\end{center}
\caption{A comparison of elemental abundances for Magellanic RGB and AGB-O stars showing that the AGB-O stars give consistent abundances for many elements.  
}
\label{fig:agbelemcomp}
\end{figure*}

\begin{figure}[t]
\begin{center}
\includegraphics[width=0.6\hsize,angle=0]{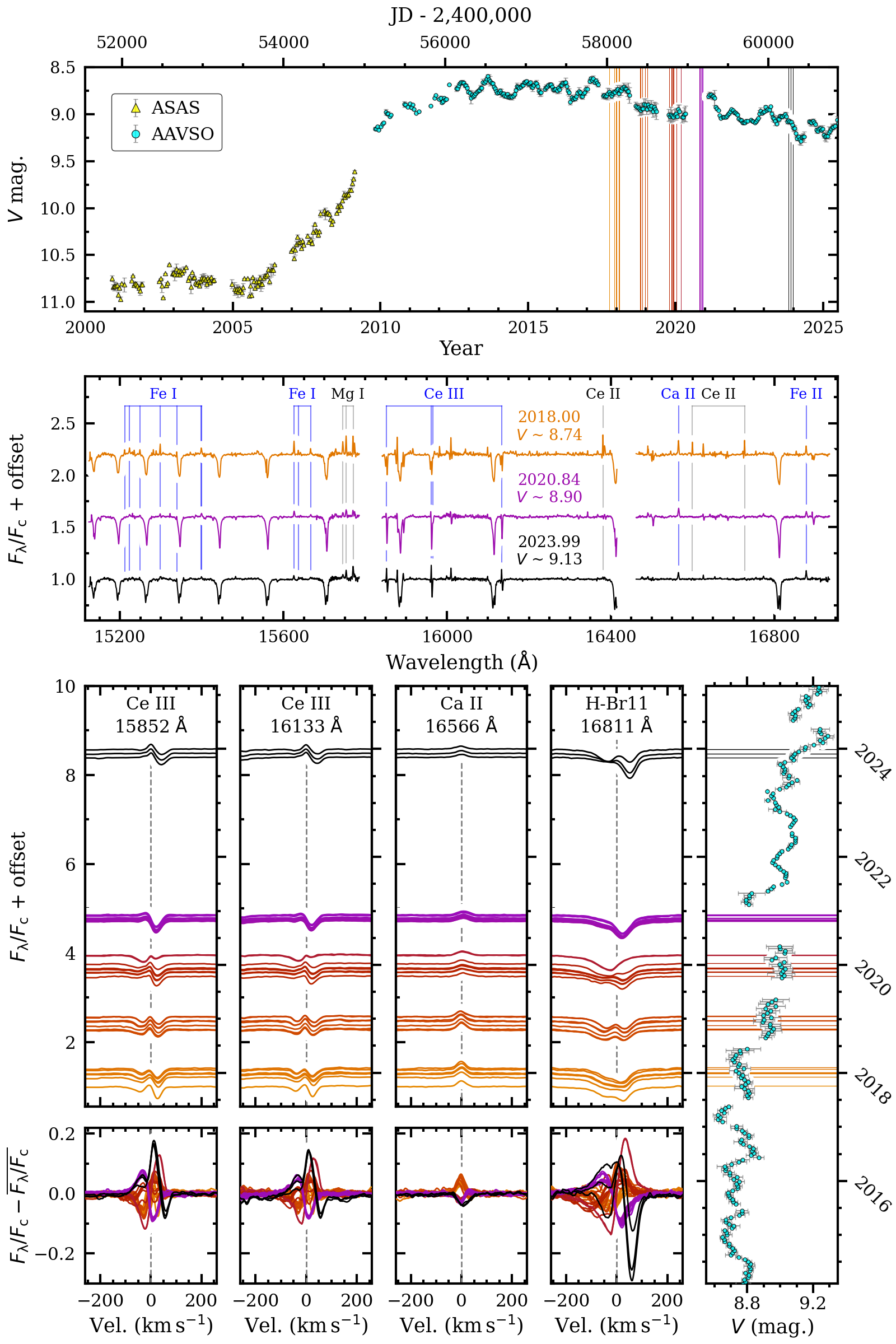}
\end{center}
\caption{Photometric and spectroscopic variability of the LMC LBV RMC 71 ($=$HD\,269006) during the decline of its dramatic, multi-decade outburst. \emph{Upper panel:} 25 years of almost continuous ASAS and AAVSO $V$-band photometry that captured the almost 2.5 magnitude brightening. Vertical lines indicate \textsl{APOGEE} observation epochs, which took place near the apparent brightness peak. \emph{Middle panel:} identification of some of the metal lines present in most of the \textsl{APOGEE} spectra. Shockingly, the strongest features throughout the spectra are Ce~{\sc iii} lines, and these are also accompanied by Ce~{\sc ii} lines in most epochs. \emph{Lower panels:} a detailed look at the perplexing variability of RMC 71 over the course of 2017 -- 2024. Whereas the hydrogen and Ce~{\sc iii} lines alternate from emission+absorption (2018--2020) to mostly absorption (late 2020) to inverse P\,Cygni (late 2023), lines from Ca~{\sc ii} and Fe~{\sc ii} maintain pure emission morphology throughout.}
\label{fig:rmc71}
\end{figure}

\begin{figure}[t]
\begin{center}
\includegraphics[width=0.6\hsize,angle=0]{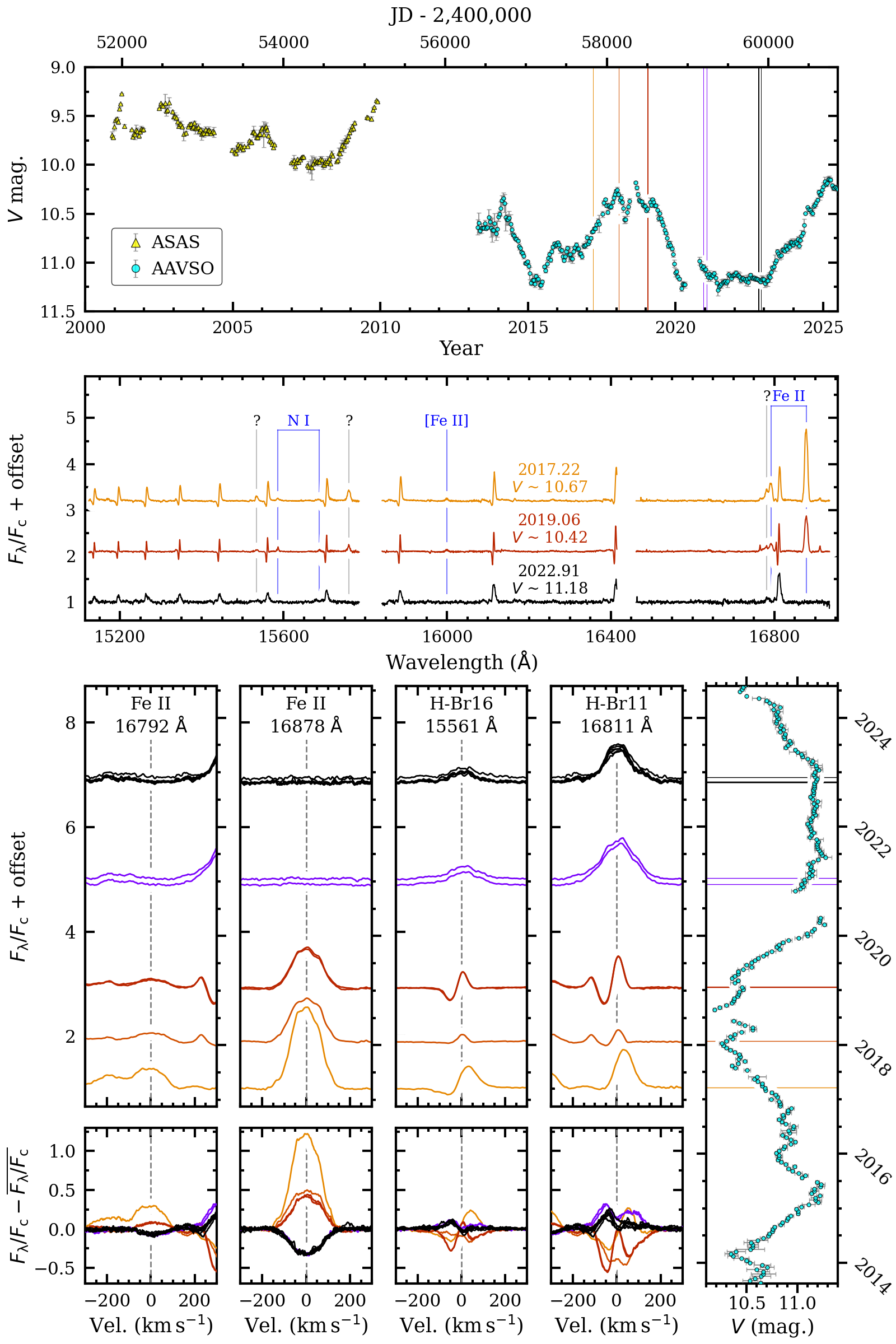}
\end{center}
\caption{Similar to Figure~\ref{fig:rmc71} but for the LMC LBV RMC 127 ($=$HD\,269858), which faded by almost 2.5 magnitudes between 2001--2021. Most of the \textsl{APOGEE} spectra were taken during relatively faint (and hence hot) phases, such that no absorption features are present save for the P\,Cygni morphology of the hydrogen lines in early 2019. The complete disappearance in the post-2020 spectra of the Fe~{\sc ii}~16878\,{\AA} emission line that was very strong in the pre-2020 spectra is among the the most extreme variability of any star observed by \textsl{APOGEE} to date. Note that some of the emission features remain unidentified and thus are labeled with question marks in the middle panel.}
\label{fig:rmc127}
\end{figure}

\begin{figure}[t]
\begin{center}
\includegraphics[width=0.7\hsize,angle=0]{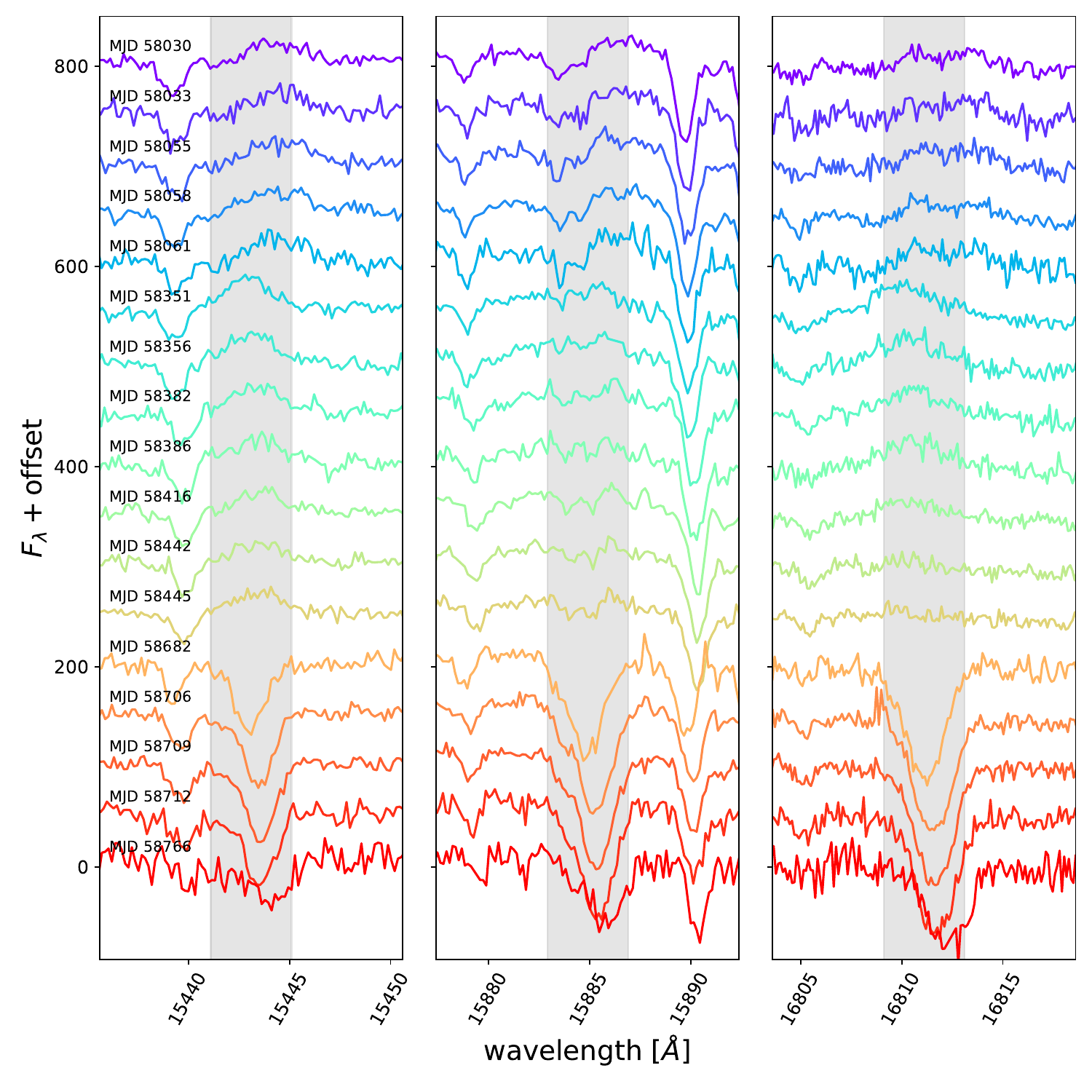}
\end{center}
\caption{Time-series \textsl{APOGEE}-2S spectra of the LIN358 symbiotic binary star system in the SMC showing large spectral variations over multiple years \citep{Washington2021}.}
\label{fig:lin358}
\end{figure}


\end{document}